\documentclass[final,5p,times,twocolumn]{elsarticle}

\usepackage{amsfonts, amsmath, amsthm, amssymb}

\usepackage{color,soul}

\usepackage{subcaption}
\usepackage{multicol} 
\usepackage{stackrel}
\usepackage{graphicx}
\usepackage{listings}
\usepackage{csquotes}
\usepackage[hidelinks]{hyperref}
\usepackage[nottoc,notlot,notlof]{tocbibind}
\usepackage{cancel}
\usepackage{xcolor}
\usepackage{braket}
\usepackage{bm} 

\begin{document}

\begin{frontmatter}

\title{Oscillations and Spatial Patterns in Large-Scale Stochastic Gene Regulatory Networks}

\author[1]{Manuel Eduardo Hernández-García}
\ead{manuel.hernandezgarcia@viep.com.mx}

\author[1]{Jorge Velázquez-Castro}
\ead{jorge.velazquezcastro@correo.buap.mx}

\address[1]{Facultad de Ciencias Físico-Matemáticas, Benemérita Universidad Autónoma de Puebla, Avenida San Claudio y 18 Sur, Col. San Manuel, Puebla, México.}

\begin{abstract}
Gene regulatory networks (GRNs) are fundamental to cellular growth and tissue formation, orchestrating spatially and temporally regulated gene expression during development. These networks are inherently subject to intrinsic fluctuations arising from molecular noise, making the analysis of their stability essential for understanding robust pattern formation and developmental dynamics of the organism. In this study, we analyze the stability and dynamics of cyclic GRNs with negative feedback and diffusion, considering both deterministic and stochastic approaches. In the deterministic case, the system exhibits a bifurcation between stability and instability, leading to Hopf instability in the absence of diffusion and to Turing-Hopf instability when diffusion is included. It was observed that the discretization of the spatial domain introduces additional unstable modes, enabling a wider range of patterns. The stochastic framework based on the second-moment approach, which incorporates intrinsic fluctuations, reveals that for small system sizes, fluctuations can dominate the dynamics and induce stochastic Turing instability, even when the system is stable in the absence of diffusion. Notably, Turing instabilities can emerge even when all variables have the same diffusion rate. The developed framework provides a systematic method for analyzing the stability of high-dimensional stochastic systems with diffusion, thereby simplifying the prediction of Turing and Turing-Hopf instabilities. These findings contribute to a deeper understanding of the complex dynamics and pattern formation in GRNs, with potential implications for biological processes, such as cellular differentiation and development.

\end{abstract}


\end{frontmatter}

\section{Introduction}

The development of an organism involves multiple cells and is a complex spatiotemporal process. The principal mechanism behind this process is gene expression, which ultimately generates patterns of differentiated cell types that characterize organisms \cite{Gaffney}. Gene regulatory networks (GRNs) are central to cellular regulation and development, allowing cells to adapt their gene expression in response to the environment \cite{Walk}. Notable instances include the p53-Mdm2 feedback loop \cite{Lev, Extri}, in which p53 functions as a tumor suppressor and mutations in its regulatory pathway are frequently associated with cancer \cite{Smeenk, Baroni}, as well as the synthesis of Hes-1, a transcription factor implicated in stem cell maintenance, cellular differentiation, and oncogenesis \cite{Hirata, Liu}.

In this study, we focused on stochastic GRNs governed by negative feedback mechanisms \cite{Perez, Schiavon}. On the other hand, within deterministic frameworks, where stochastic effects are disregarded, such networks are known to exhibit oscillatory dynamics, particularly when delays or intermediary components are present \cite{Xiao, Takada}. Cyclic behavior and oscillations are integral to various biological processes, with circadian rhythms serving as prominent examples \cite{Vitaterna, Mills}. 

Oscillations in GRNs with negative regulation can arise in this type of system by the Hopf instability. A Turing instability indicates whether a system can generate spatial patterns \cite{Turing}, and it has also been applied to systems that involve gene expression \cite{Diambra}. When spatial diffusion is introduced, if the system remains unstable and both the Turing and Hopf conditions are simultaneously satisfied, a Turing–Hopf instability occurs. In this case, spatial patterns not only emerge but also oscillate in time, giving rise to dynamic, wave-like structures. Mathematical modeling demonstrates that diffusion coefficients act as bifurcation parameters, giving rise to Hopf bifurcations in systems like the Hes-1 transcription factor network \cite{Chaplain}.

Nevertheless, many regulatory networks function under conditions of low molecular abundance, making them susceptible to intrinsic stochastic fluctuations. Extrinsic fluctuations can also be considered; they are expressed as fluctuations in the model parameters and represent fluctuations of environmental factors \cite{Extri, Thomas}. However, in this study we focus exclusively on intrinsic fluctuations and on diffusion, which arise from the discrete and probabilistic nature of molecular interactions \cite{Alon, VecchioM, Gar}.

Several approaches have been developed to analyze stochastic systems, including the chemical master equation \cite{Gar}, stochastic simulation algorithms such as Gillespie’s method \cite{Gillespie}, and continuum approximations such as the Langevin equation \cite{VecchioM, Langevin}, Fokker–Planck equations \cite{Scott} and the linear noise approximation \cite{Lot, Grimagene}. An alternative and widely used method is the moment-based approach \cite{Exact, Gomez, Manuel}, particularly the second-moment approximation, which captures the system dynamics through a set of ordinary differential equations (ODEs) describing the time evolution of the means and second central moments. To account for nonlinearities, such as those introduced by Hill-type kinetics, higher-order moments may be incorporated \cite{Lakatos}. Remarkably, in Hernández et al. \cite{Exact} it was shown that systems involving only first-order reactions combined with nonlinear Hill functions can be described exactly using this approach. This ODEs-based representation facilitates the use of dynamical systems tools, including stability and bifurcation analyses \cite{Stability}, to characterize behaviors such as steady states and periodic orbits \cite{Smith}.

The present study aims to examine large-scale GRNs with negative feedback and diffusion under conditions dominated solely by intrinsic fluctuations \cite{Kulasiri, Ribeiro} and to identify the conditions under which these systems exhibit dynamical instabilities. In the absence of diffusion, oscillations can be sustained, whereas with diffusion they can exhibit Turing instabilities or Turing–Hopf instabilities. Although analyzing high-dimensional networks poses significant challenges because of the complexity of deriving instability criteria \cite{Piskovsky}, previous studies have addressed such issues in deterministic settings using control-theoretic techniques \cite{Harat, Hori, Hori2}. Because the moment-based formulation converts stochastic dynamics into a set of ODEs, these analytical tools can be extended to study stochastic systems, as in the work of \cite{Oscillations}. 

Stochastic Turing instability represents a significant extension of classical pattern formation theory, in which intrinsic fluctuations enable spatial pattern formation beyond the traditional deterministic parameter regime. Multiple studies have demonstrated that stochastic fluctuations can induce Turing-like patterns in parameter regions where deterministic analysis predicts no instability  \cite{Biancalani}. Because we are considering a stochastic model, we also explore whether this type of instability appears in the system. 

To validate our approach, we considered a representative example: the repressilator, a synthetic genetic oscillator initially engineered by Elowitz et al. \cite{Elowitz} and studied in deterministic frameworks \cite{Hanna} and, in some cases, with diffusion \cite{Macnamara}. These case studies illustrate the scalability of our methodology and reveal how the system size and diffusion affect the stability of the system in both deterministic and stochastic approaches.

The remainder of this paper is organized as follows: In Section \ref{section2}, we present the set of ODEs describing the mean and second central moments with diffusion. In Section \ref{section3}, we analyze a cyclic GRN with negative feedback and the effects of diffusion on stability in both deterministic and stochastic frameworks. In Section \ref{section4} we analyze the repressilator in one dimension, also in deterministic and stochastic approaches. In Section \ref{section5}, we analyze a similar system in two dimensions to study the stability and show the patterns that result from the solution. Finally, in Section \ref{section6}, we summarize the results and conclusions.

\section{Moment Approach} \label{section2}  

In this section, we present a system of ordinary differential equations (ODEs) describing the evolution of the mean concentrations and their second central moments, including diffusion. These equations are derived from the reaction--diffusion master equation (RDME), which provides a fundamental framework for modeling stochastic chemical systems. In this study, we restrict our analysis to intrinsic fluctuations. Accordingly, we follow the methodology developed in \cite{Gar} and directly present the resulting ODEs. The detailed derivations can be found in \cite{Exact}. We also introduce the necessary definitions that formalize the concept of a chemical reaction--diffusion network. We begin by defining the chemical reaction network. \\   

\textbf{Definition 1 \cite{Anderson}.} A chemical reaction network is a triplet of non-empty, finite sets, usually denoted by $\mathcal{N}= \{\mathcal{S}, \mathcal{C}, \mathcal{R}\}$, where:

\begin{enumerate}     

\item A set of $N$  chemical species denoted by $\mathcal{S}= \{\mathcal{S}_1, \mathcal{S}_2, ..., \mathcal{S}_N\}$.      

\item A set of non-negative integer linear combinations of the species denoted by    

{\footnotesize
$\mathcal{C}=\{ \sum_{l=1}^{N} \alpha_{1l} \mathcal{S}_l, \sum_{l=1}^{N} \alpha_{2l} \mathcal{S}_l, ..., \sum_{l=1}^{N} \alpha_{nl} \mathcal{S}_l, $}\hfill \\
 \hspace*{24ex} {\footnotesize 
$\sum_{l=1}^{N} \beta_{1l} \mathcal{S}_l, \sum_{l=1}^{N} \beta_{2l} \mathcal{S}_l, ..., \sum_{l=1}^{N} \beta_{nl} \mathcal{S}_l \}$}, \\

($i=(1,2,...,m)$, $l=(1,2,...,N)$) coefficients $\alpha_{il}$ and $\beta_{il}$ are non-negative integers, and they represent the stoichiometric coefficients.  Additionally, the stoichiometric matrix is defined as $\Gamma_{li}= \beta_{li}-\alpha_{li}$, where the exchange of indices denotes transpose.  

\item A set of $m$  chemical reactions denoted by $\mathcal{R}= \{\mathcal{R}_1, \mathcal{R}_2, ..., \mathcal{R}_m\}$. Through which these species are transformed, represented as 
\begin{align}      
\mathcal{R}_i : \sum_{l=1}^{N} \alpha_{il} \mathcal{S}_l \stackbin[]{k_i}{\rightarrow} \sum_{l=1}^{N} \beta_{il} \mathcal{S}_l, \label{1}
\end{align}
where $k_i$ is a kinetic parameter that denotes the rate of the reactions. The order of a reaction $\mathcal{R}_i$ is defined by $O(\mathcal{R}_i)=\sum_l \alpha_{il}$.

\end{enumerate}  
 \begin{figure} [h!t]  
 \centering 
 \includegraphics[width=0.24\textwidth]{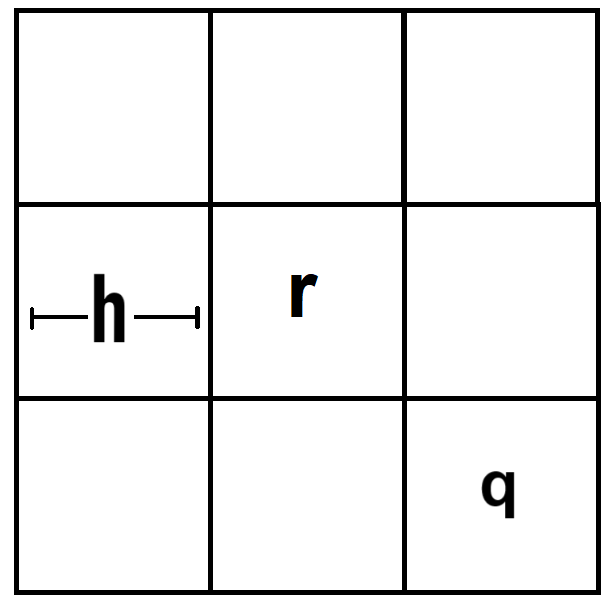}   \caption{ \textbf{Space domain partitioned into voxels.} The spatial domain is partitioned into uniformly sized voxels with side length $h$. Diffusive transport of chemical species is permitted between any pair of voxels $r$ and $q$, not limited to adjacent ones.} \label{fig.1} 
 \end{figure}   

If the reactions take place in an extended spatial domain, it is partitioned into voxels for convenience. The voxels are characterized with a side length $h$ \cite{Exact, Lot}, as shown in Figure \ref{fig.1}. Thus, we define a chemical reaction diffusion network. \\

 \textbf{Definition 2 \cite{Exact}.} A chemical reaction-diffusion network is a non-empty set \(\mathcal{D}= \{ \mathcal{J}, \mathcal{N}^d \}\), where: 

 \begin{itemize}     
 \item \(\mathcal{J} \subset \mathbb{R}^3\) is a spatial domain discretized into \(J\) voxels of uniform characteristic size $h$, adopting a geometry like the shown in Fig. \ref{fig.1}. The voxels do not overlap and cover the entire domain.

 \item Within each voxel \(r\) of \(\mathcal{J}\) (\(r \in \{1, 2, \dots, J\}\)), there exists a chemical reaction network, \(\mathcal{N}^r = \{\mathcal{S}^r, \mathcal{C}^r, \mathcal{R}^r\}\), where:    

 \begin{itemize}        
 \item \(\mathcal{S}^r\): the set of $N$ species in voxel \(r\), 
 \item \(\mathcal{C}^r\): the set of complexes in voxel \(r\),      
 \item \(\mathcal{R}^r\): the set of $n$ reactions in voxel \(r\).     
 \end{itemize} 

 From this, the set of chemical reaction networks over \(\mathcal{J}\) is denoted as \(\mathcal{N}^d = \{\mathcal{N}^1, \mathcal{N}^2, \dots, \mathcal{N}^J\}\).

 \item Let \(\mathcal{S}^r\) be a set of \(N\) chemical species in voxel \(r\),  which elements subsequently move to voxel \(q\). The diffusion process is expressed as follows,    
 \begin{align}         
 \mathcal{S}_{l}^r  \stackbin[]{d_{rq}^l}{\rightarrow} \mathcal{S}_{l}^q, \label{2}      
 \end{align} 
 where \(d_{rq}^l\) denotes the diffusion rate of the species \(\mathcal{S}_l^r\) ($\in \mathcal{S}^r$) from voxel \(r\) to voxel \(q\) (molecules per unit time) and $d_{rr}^l =0$. 
 \end{itemize}

From the previous definition, the RDME can be derived \cite{Exact}. For convenience, we present here the ODEs obtained from it that describe the dynamics of the mean concentration in each voxel $r$, $s_l^r$ and the second central moment between chemical species from two voxels $r_1$ and $r_2$  $M^2_{l_1^{r_1},l_2^{r_2}}$ ($l_1,l_2=(1,2,...,N)$). These equations are obtained directly from the RDME. According to the result in Hernandez et al. \cite{Exact}(Corollary 1), when the system involves only zero- and first-order reactions and in the reaction rates are functional parameters \( k_i = k_i^* f_i(\mathbf{s}, \mathbf{M}^2) \), where each function \( f_i \) depends on the mean concentrations and the second central moments, such as Hill functions commonly used in gene regulatory networks, the set of ODEs in \eqref{3} describes exactly the dynamics of the system. However, if the functional parameters involve dependencies on higher-order moments, a more general formulation is required to accurately capture the system behavior \cite{Exact}. \\

\textbf{ Proposition 1 \cite{Exact}. \label{P.1}} Let \(\mathcal{D}\) be a chemical reaction-diffusion network, where:

\begin{itemize}
    \item There are only zero- and first-order reactions in each voxel \(r\).

    \item The parameters have a functional form that depends on the mean concentrations $\mathbf{s}^r$ and second central moments \(\mathbf{M}_{r,r}^2\), such that \[     k_i^r = k_i^{*r} f_i(\mathbf{s}^r, \mathbf{M}^2_{r,r}).   \] 
\end{itemize}

The system can then be described exactly using the following ODEs: 
\begin{subequations}\label{3}
{\footnotesize
\begin{align}     
\frac{\partial s_l^r }{\partial{t}} & =  \sum_{i} \Gamma_{li} k_i^r R_i(\mathbf{s}^r)   +  \sum_q  \left(  d_{qr}^{l} {s_l^q} - d_{rq}^{l} {s_l^r} \right),  \\
\frac{\partial M^2_{{l_1}^{r_1}, {l_2}^{r_2}}}{\partial{t}}  & = \sum_{i} \left( \delta_{r_1,r_2} \frac{\Gamma_{l_1 i} \Gamma_{l_2 i}}{\Omega} \left(k_i^{r_1} R_i(\mathbf{s}^{r_1})   \right) \right. \nonumber \\
&+ \left. \sum_{j_1=1}^{N} \left( M^2_{{l_1}^{r_1}, {j_1}^{r_2}} \Gamma_{l_2 i} k_i^{r_2} \frac{\partial R_i(\mathbf{s}^{r_2})}{\partial {s_{j_1}^{r_2}}}  + M^2_{{j_1}^{r_1}, {l_2}^{r_2}} \Gamma_{l_1 i} k_i^{r_1} \frac{\partial R_i(\mathbf{s}^{r_1})}{\partial {s_{j_1}^{r_1}}} \right)   \right) \nonumber \\
&+  \frac{\delta_{l_1, l_2}}{\Omega} \left( \delta_{r_1,r_2} \sum_q \left( d_{q r_1}^{l_1} s_{l_1}^{q} + d_{r_1 q}^{l_1} s_{l_1}^{r_1} \right) -(d_{r_1 r_2}^{l_1} s_{l_1}^{r_1} + d_{r_2 r_1}^{l_1} s_{l_1}^{r_2}) \right) \nonumber \\
&+  \sum_q \left( d_{qr_1}^{l_1} M^2_{l_1^{q},l_2^{r_2}}  - d_{r_1q}^{l_1} M^2_{l_1^{r_1},l_2^{r_2}} \right) +  \sum_q \left( d_{qr_2}^{l_2} M^2_{l_1^{r_1},l_2^{q}}  -d_{r_2q}^{l_2} M^2_{l_1^{r_1},l_2^{r_2}}  \right),  
\end{align}}
\end{subequations}

where $\mathbf{s}^r=(s_1^r,s_2^r,..., s_N^r)$  and $R_i(\mathbf{s}^{r})= \prod_j (s_j^r)^{\alpha_{ij}}$. \\

In this work, we consider the system in a discretized space; however, in the limit in which $h$ tends to zero, it is recovered the continuous case \cite{Lot}. 

We consider that discretization of space is not always an approximation but, at some scales, better describes the behavior of biological tissues, where the discrete nature may account for the properties of tissues composed of discrete cells.

\section{Gene Regulatory Network with Negative Feedback} \label{section3} 

\begin{figure}[h!t]
    \centering
    \includegraphics[width=0.8\linewidth]{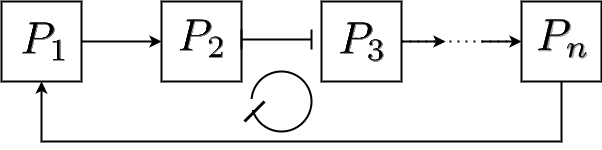}
    \caption{\textbf{Cycle gene regulatory network.} This figure shows a cyclic gene regulatory network with negative feedback, where the arrows represent positive regulation, and the bars represent negative regulation. The complete cycle has a negative feedback, and in each node, there is a module of transcription-translation processes that synthesizes a protein that acts as a transcription factor for the subsequent module. }
    \label{fig.2}
\end{figure}

Gene regulatory networks (GRNs) are essential for the organization of living systems because they govern the key mechanisms of cellular regulation and development. Through these networks, cells can adapt their gene expression patterns in response to environmental conditions \cite{Walk}. A particularly relevant property arises in GRNs endowed with negative feedback: under deterministic descriptions, such systems can generate sustained oscillations, provided that regulatory delays or intermediate interactions are present \cite{Xiao, Takada}. Previous investigations have demonstrated that within a stochastic framework, the system size $\Omega$ also acts as a critical parameter that controls the onset of oscillatory behavior \cite{Oscillations}. In this study, we aimed to determine whether diffusion-driven stochasticity could also sustain oscillations in these systems.

Oscillatory mechanisms of this type have been reported in several biological contexts. A prominent case is the regulatory feedback between p53 and Mdm2 \cite{Lev, Extri}, where p53 functions as a tumor suppressor that is frequently altered during carcinogenesis \cite{Smeenk, Baroni}. Another well-studied example is the regulation of Hes-1, a transcription factor indispensable for stem cell maintenance, cellular fate decisions, and tumor development \cite{Hirata, Liu}. However, this type of system has a spatial domain, then its important to consider the effects of the diffusion as in the embryonic stem cells \cite{Sturrock, Alana}. 

Our analysis focuses on the cyclic GRN shown in Fig. \ref{fig.2}. Each node in the cycle represents a transcription–translation process that produces a protein that acts as a transcription factor. These factors control other nodes through Hill-type interactions, either activating or repressing them, and together, they give rise to an effective, negative feedback loop. The topology, first suggested in \cite{Banks}, is generalized here to include the influence of diffusion and intrinsic stochastic fluctuations. This extends previous approaches \cite{Harat, Hori, Hori2} by incorporating stochastic corrections into the Hill functions and diffusion in a discrete spatial domain, following the methodology developed in \cite{Exact}.

\subsection{Model}

We examined the cyclic gene regulatory network depicted in Fig. \ref{fig.2}, where regulatory links are represented by arrows for activation and bars for repression, highlighting the overall negative feedback organization across $n$ nodes. Since each transcription–translation unit follows analogous reaction schemes, the reactions within an arbitrary module $j$ ($j \in (1,2,...,n)$) are

\begin{table}[h!]
    \centering
    \begin{tabular}{|c|c|}
        $ \emptyset \stackbin{\gamma_{1,j}}{\longrightarrow} M_j$ & $M_j  \stackbin{\delta_{1}}{\longrightarrow} \emptyset$ \\
        $M_j    \stackbin{\gamma_{2,j}}{\longrightarrow} M_j + P_j  $ & $ P_{j}  \stackbin{\delta_{2}}{\longrightarrow} \emptyset $ \\
    \end{tabular}
\end{table}
the first reaction is the synthesis of the mRNA by a functional parameter $\gamma_{1,j}$ (in this case, a Hill function) and its degradation by a parameter $\delta_1$, the next reaction is the synthesis of the proteins by a parameter $\gamma_{2,j}$   and its degradation by a parameter $\delta_2$.  The degradation parameters of the mRNA and proteins were the same for all modules.  In addition, diffusion occurs only for proteins because they move between voxels that represent cells, whereas mRNA is in cells and does not diffuse \cite{Macnamara}. Then the diffusion processes are
\begin{align}
     P_{j}^r  \stackbin{D}{\longrightarrow}  P_{j}^q,
\end{align}
were for the sake of simplicity we have assumed that all proteins have the same diffusion parameter. The stoichiometric coefficients in the module $j$ are $\alpha_{il}^j$ and $\beta_{il}^j$, and the stoichiometric matrix for each module are,

{\small
\begin{align}
 \alpha_{il}^j&= \begin{pmatrix}
		0& 0   \\
            1& 0   \\
            1& 0   \\
            0& 1   \\
	\end{pmatrix} ,    &
 \beta_{il}^j&=  \begin{pmatrix}
		1& 0   \\
            0& 0   \\
            1& 1   \\
            0& 0   \\
	\end{pmatrix},  &
 \Gamma_{li}^j&= \begin{pmatrix}
		1& -1 & 0 & 0   \\
            0 & 0 & 1& -1      \\
	\end{pmatrix}, \nonumber
\end{align}}
and,
\begin{align}
   R_{1,j}&= 1  , &
    R_{2,j}&=  m_j, \nonumber \\
   R_{3,j}&= m_j   , &
    R_{4,j}&=  p_j,
\end{align}
where $m_j$ and $p_j$ are the mean concentrations of mRNA and protein of module $j$. If the system has $n$ nodes, then to describe the system, it is necessary to have $2n + 5$ parameters. However, if we make the system dimensionless (see \ref{B} for more details), we can reduce the number of parameters to $n+2$. Then, without loss of generality, we chose $D=1$, $\gamma_{2,j}=\delta_2=1,$ and $\delta_1=\delta$ and $\gamma_{1,j}= \delta \gamma_j^2 H_j$, where $H_j$ is the Hill function defined according to the deterministic or stochastic cases.

\subsection{Deterministic Case} \label{section3.2}

We began our analysis by considering the system within a deterministic framework (in the absence of fluctuations). As reported in \cite{Krause}, a Turing instability criterion predicts spatial patterns, but the direct numerical solution of the equations may not show them \cite{Nava}. This discrepancy may stem from the assumption of a continuous spatial domain within the analytical study, which potentially fails to account for the discrete nature of cells and molecules.

To this end, we compared the results of pattern formation obtained from both continuous and discrete one-dimensional (1D) models, emphasizing how discretization modifies the model outcomes. Furthermore, in systems with a large number of interacting chemical species \cite{Piskovsky}, the classical Turing instability criterion becomes increasingly difficult to apply because establishing stability conditions from Jacobian and diffusion matrices is not straightforward. To overcome this limitation, we adopted the generalized instability criterion proposed in \cite{Harat, Hori, Oscillations}, which was specifically designed for high-dimensional systems. We further extended this approach by incorporating diffusion into the model.

First, the Hill functions needed to describe the system in a deterministic approach are:
\begin{align}
    H_j= H_j(p_{j-1} )=  \begin{cases}         
 \left(  \frac{1}{1+ p^2_{j-1} }\right) & \text{for a repressor } \\       
 \left(  \frac{p^2_{j-1}}{1+ p^2_{j-1} }\right) & \text{for an activator}    
\end{cases}  . 
\end{align}
These are the Hill functions for a repressor or activator, respectively, in which we used a Hill coefficient equal to two and a dissociation constant equal to one. The equations for the  concentrations considering diffusion are as follows:
\begin{subequations}\label{7}
\begin{align} 
    \frac{\partial m_j^r}{ \partial t}=& \delta( \gamma_j^2 H_j - m_j^r),  \\
    \frac{\partial p_j^r}{ \partial t}=&  m_j^r - p_j^r + \frac{p_j^{r+h}+p_j^{r-h} - 2p_j^r}{h^2} . 
\end{align}
\end{subequations}
Note that these equations are only in 1D. We used the fact that the proteins only moved to their first neighbors. To make a stability analysis of this system, we represent the solution as a discrete Fourier transform
\begin{subequations} \label{8}
\begin{align}  
   m_j^K= \mathcal{F}[m_j^r]= \sum_r m_j^re^{-IKr}, \\
    p_j^K=\mathcal{F}[p_j^r]= \sum_r p_j^re^{-IKr},
\end{align}
\end{subequations}
($I$ is the complex unit) then, it follows that 
\begin{align}
   \mathcal{F}\left[ \frac{p_j^{r+h}+p_j^{r-h} - 2p_j^r}{h^2}  \right]=& - \left( \frac{sin\left( K \frac{h}{2} \right)}{\frac{h}{2}}\right)^2 p_j^K \nonumber \\
   =&-f_1(K,h) p_j^K . \label{9}
\end{align}
In the limit in which $\lim_{h \to 0} f_1(K,h)  = K^2$, that is the expected behavior when the space is considered continuous\footnote[1]{In three dimensions, the function is $f_1(K_1,K_2,K_3,h)= \left( \frac{sin\left( K_1 \frac{h}{2} \right)}{\frac{h}{2}}\right)^2 + \left( \frac{sin\left( K_2 \frac{h}{2} \right)}{\frac{h}{2}}\right)^2 + \left( \frac{sin\left( K_3 \frac{h}{2} \right)}{\frac{h}{2}}\right)^2$, and the limit, $\lim_{h \to 0} f_1(K_1,K_2,K_3,h)  = K_1^2 + K_2^2 + K_3^2$.}. Considering Eq. \eqref{9} in Eq. \eqref{7} we get
\begin{subequations} \label{10}
\begin{align}
    \frac{\partial m_j^K}{ \partial t}=& \delta( \gamma_j^2 \hat{H}_j - m_j^K),  \\
    \frac{\partial p_j^K}{ \partial t}=&  m_j^K -(1+f_1(K,h)) p_j^K. 
\end{align}
\end{subequations}
where $\hat{H}_j= \sum_r \frac{1}{1+ (p_j^r)^2} e^{-IKr}$. 
In the absence of diffusion, the stationary state of the system is 
\begin{align}
    p_{j,ss}= m_{j,ss}, 
\end{align}
and to obtain the values of the stationary concentrations of the mRNA, we need to solve the following equations
\begin{align}
    m_{j,ss} = \begin{cases}        
    \gamma_j^2 \left(  \frac{1}{1+ m^2_{j-1,ss}  }\right) & \text{for a repressor } \\  
 \gamma_j^2 \left(  \frac{m^2_{j-1,ss} }{1+ m^2_{j-1,ss}  }\right) & \text{for an activator}    
\end{cases}. \label{12}
\end{align}

In the methodology introduced by \cite{Harat, Hori}, stability analysis requires linearization of the system around its stationary state, and then a transfer function is employed to assess stability. For GRNs with negative feedback, the dynamics may lead either to a stationary point or to a limit cycle, the latter corresponding to an unstable regime. To assess stability, we follow a similar approach as in \cite{Harat, Hori}. The first essential component is the derivative of the Hill function evaluated at the steady state,
\begin{align}
    \Xi^D_j= & \left. \frac{\partial  H_j}{ \partial p_{j-1}} \right|_{ss} = \left(  \frac{2 m_{j-1,ss}}{ 1+ m^2_{j-1,ss}}\right)\times\begin{cases}         
  -1 & \text{for a repressor } \\  
 +1 & \text{for an activator}  \end{cases}.
\end{align}
The following variables are introduced to carry out the linearization; $\Delta m_j^K= m_j^K - m_{j, ss}^K$, $\Delta p_j^K= p_j^K - p_{j, ss}^K$. Then for each module, its previous transcription–translation module is treated as an applied control, $u_j$. Thus we write the equations for module $j$ as 

\begin{figure}[h!t]
    \centering
    \includegraphics[width=0.47\linewidth]{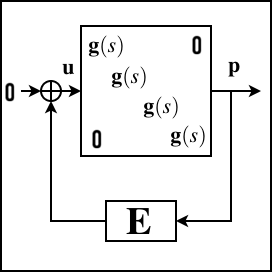}
    \includegraphics[width=0.47\linewidth]{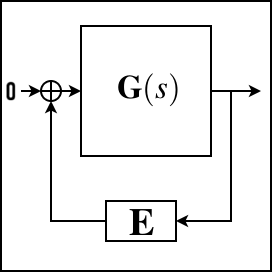}
    \caption{\textbf{Block diagram.} On the left, we show the overall GRN, composed of $n$ identical modules, each with internal dynamics given by $\mathbf{g}(s)$, an input $u$, and outputs $p$. On the right panel, we present a modified GRN that is equivalent to the one on the left, describing the interconnection of the same $n$ identical modules by $\mathbf{G}(s)$, which has $n$ inputs and $n$ outputs.} 
    \label{fig.1a}
\end{figure}

\begin{align}
\frac{\partial}{\partial t}
\begin{pmatrix}
\Delta m_j^K \\
\Delta p_j^K 
\end{pmatrix}
=&
\begin{pmatrix}
-\delta & 0  \\
1 & -(1+f_1(K,h)) 
\end{pmatrix}
\begin{pmatrix}
\Delta m_j^K \\
\Delta p_j^K 
\end{pmatrix}
+ 
\begin{pmatrix}
\delta \\
0 
\end{pmatrix} u_j  . \label{linear}
\end{align}
These equations are written in the form $\frac{\partial}{\partial t} \mathbf{x}_j= \mathbf{A} \mathbf{x}_j + \mathbf{B} u_j$ with $u_j=\gamma_j^2( \Xi^D_j \Delta p_{j-1}^K )$. Thus, Eq. \eqref{linear} gives $\mathbf{A}$ and $\mathbf{B}$. Furthermore, $\mathbf{A}$ and $\mathbf{B}$ are similar for all modules. 

Now we write $u_j$ in a matrix form $\mathbf{u}= \mathbf{E} \Delta\mathbf{p}^K$

{\small
\begin{align}
\begin{pmatrix}
u_1 \\
u_2 \\
u_3 \\
\vdots\\
u_n
\end{pmatrix}
= 
\begin{pmatrix}
0 & 0 & \dots & 0& \gamma_1^2 \Xi^D_1  \\
\gamma_2^2\Xi^D_2  & 0 &  \dots& 0 &0\\
0 & \gamma_3^2 \Xi^D_3  &  \dots &0 &0  \\
\vdots &  \vdots & \ddots &\vdots & \vdots \\
 0 &0 & \dots& \gamma_n^2 \Xi^D_n  & 0 \\
\end{pmatrix}
\begin{pmatrix}
\Delta p_1^K \\
\Delta p_2^K \\
\vdots \\
\Delta p_{n-1}^K \\
\Delta p_n^K \\
\end{pmatrix} \label{matrix}
\end{align}}

From $\mathbf{E}$, we calculate its eigenvalues, 
\begin{align}
   \lambda_j^D= L^D e^{I(2j-1)\pi/n},  L^D= \prod_{j=1}^n \left| \gamma_j^2 \Xi^D_j \right|^{1/n}.
\end{align}
The protein  of a module $j$ affects the next module in the cycle, and then its values are the output in each module; thus, we write
\begin{align}
 \mathbf{y}_j=   \mathbf{Cx}_j= 
    \begin{pmatrix}
0 & 1  \\
\end{pmatrix}
\begin{pmatrix}
\Delta m_j^K \\
\Delta p_j^K 
\end{pmatrix},
\end{align}
matrix $\mathbf{C}$ is similar for all modules. Following the procedure to determine the stability conditions of the controllable variables \cite{Harat, Hori}, we now calculate the transfer function of each module,
\begin{align}
    \mathbf{g}(s)=& \mathbf{C (}s\mathbf{I - A)^{-1} B} \nonumber \\
    =& \frac{\delta}{(\delta + s)(1+f_1(K,h)+s)} =  g_1(s).
\end{align}

We can now write the transfer function of the overall GNR (see Fig. \ref{fig.1a}). First, we define
\begin{align}
    \mathbf{G}(s)= 
g_1(s) I_{n\times n} ,
\end{align}
Then, the overall transfer function of the system is  
\begin{align}
    \mathcal{H}(s)=&  (\mathbf{I} - \mathbf{G}(s)\mathbf{E})^{-1} \mathbf{G}(s) \nonumber \\
    =& g_1(s) (I_{n \times n} - g_1(s) \mathbf{E})^{-1} \nonumber \\
    =& (\phi_d(s) I_{n \times n}- \mathbf{E})^{-1}
\end{align}
where $\phi_d(s)= 1/g_1(s) $. By finding the poles of this expression, we can determine the stability of the system. We summarize the stability condition with the following Proposition.

\newpage
\textbf{Proposition 2.} Consider a gene regulatory network system modeled by Eqs. \eqref{7}, 

\begin{itemize}
    \item[i)] If there is at least one eingenvalue $\lambda_j^D$ within the domain $\Lambda^+$, defined by $\Lambda^+= \phi_d(\mathbb{C}_+)$, where $\mathbb{C}_+= \{  s \in \mathbb{C}_{+}| Re(s) > 0  \}$ are the complex with non-negative real part, then the system is unstable.

    \item[ii)] Otherwise, the system is stable. 
\end{itemize}

In the study by Hori et al. \cite{Hori}, stability is obtained from the transfer function, which specifies the regions where the system remains stable or becomes unstable. Here, we extend this idea to incorporate the effects of diffusion in the system by making a similar analysis for each Fourier mode. When diffusion is absent, concentration oscillations emerge if the system is unstable. However, with diffusion, instability may also arise from certain values of $K$ and $h$, which are associated with pattern formation and depend on the voxel size.  From this scenario, an analytical criterion can be derived. To this end, we first define the following quantities: the ratio between the product of degradation rates and their mean,
\begin{align}
    Q= \frac{2 \sqrt{\delta(1+ f_1(K,h))} }{\delta + 1 + f_1(K,h) }. \label{15}
\end{align}
Next, we compute a parameter that helps identify the region where instability arises,
\begin{align}
    W(n,Q)= \frac{2 \left( -cos\left( \frac{\pi}{n} \right) + \sqrt{cos^2\left( \frac{\pi}{n}\right) + Q^2 sin^2\left( \frac{\pi}{n} \right) } \right) }{Q^2 sin^2\left( \frac{\pi}{n} \right)}.
\end{align}

The previous expression is based on the analytical criterion given by \cite{Hori}, which established the conditions of stability. It is worth noting that in the previous expression, the contribution of diffusion was incorporated by the function $f_1(K,h)$ in Eq. \eqref{15}. Then the following proposition follows: \\ 

\textbf{Proposition 3.} Consider a gene regulatory network system modeled by Eqs. \eqref{7}, then the system is:

\begin{itemize}
    \item[i)] Unstable: if $L^D>W(n,Q)$ for any value of $n$ and $Q$.

    \item[ii)] Stable: if $L^D\leq W(n,Q)$ for any value of $n$ and $Q$.
\end{itemize}

 In conventional instability analyses, space is typically treated as continuous; however, in biological systems, cells and molecules are finite entities rather than infinitesimal quantities. Consequently, a discrete spatial description can be more appropriate than an approximation. Therefore, we will compare whether the stability conditions are modified when analyzed using continuous or discrete spatial representations.

However, one question remains: how can we identify which type of dynamics will emerge and under what conditions patterns will appear in the system? To address this issue, the following definition is given. \\

\textbf{Definition 3.} A chemical reaction diffusion network can exhibit the following behaviors:

\begin{itemize}
    \item[i)] If the system is stable in the absence of diffusion and remains stable when diffusion is included, no spatial patterns or oscillations arise (no Turing instability).

    \item[ii)] If the system is stable in the absence of diffusion but becomes unstable when diffusion is included, spatial patterns emerge (Turing instability).

    \item[iii)] If the system is unstable in the absence of diffusion but becomes stable when diffusion is included, oscillations are suppressed by diffusion (no Turing–Hopf instability).

    \item[iv)] If the system is unstable both without and with diffusion, spatial patterns and oscillations coexist (Turing–Hopf instability).
\end{itemize} 

We analyze GRNs with negative feedback, and then we can derive what type of behavior can be given to the system in the deterministic framework, which is captured by the next proposition. \\

\textbf{Proposition 4.} Consider a gene regulatory network described by Eqs. \eqref{7}. 

\begin{itemize}
    \item[a)] If $\delta \leq 1$, there can be only a Turing-Hopf bifurcation.
    \item[b)] If $\delta > 1$, there can be either Turing instability or Turing-Hopf instability.
\end{itemize}

Within the deterministic framework, Proposition 4 allows us to classify the behavior of the system according to the value of $\delta$. In particular, the emergence of Turing instability is only possible in the regime $\delta>1$. This condition is especially relevant from a biological perspective because $\delta$ is related to the degradation rates of mRNA. Because mRNA typically degrades much faster than proteins in biological systems \cite{Alon}, the biologically realistic regime corresponds precisely to $\delta>1$. Therefore, the parameter region in which Turing patterns emerge is both mathematically admissible and biologically relevant.
This is significant because it provides a mechanism for spontaneous pattern formation and the development of spatial structures, such as tissues and other organized biological arrangements. A detailed proof of Proposition 4 is provided in \ref{C}.  However, deterministic models neglect the effects of molecular fluctuations, which are unavoidable in biological systems. To construct a more realistic model, we performed a similar analysis of a stochastic system by incorporating intrinsic fluctuations.

\subsection{Stochastic Case}

The general procedure is to apply the stability condition \cite{Oscillations} to the set of ODEs that describe the system dynamics of mean concentrations and second moments. We also use a Hill-function generalization that is useful for including information about the stochastic nature of the process \cite{Exact}
\begin{align}
    H_j= H_j(p_{j-1},M^2_{p_{j-1},p_{j-1} } )=  \begin{cases}         
 \left(  \frac{1}{1+ p^2_{j-1} + M^2_{p_{j-1},p_{j-1} } - \frac{p_{j-1}}{\Omega} }\right) & \text{for a repressor } \\       
 \left(  \frac{p^2_{j-1} + M^2_{p_{j-1},p_{j-1} }  - \frac{p_{j-1}}{\Omega}}{1+ p^2_{j-1} + M^2_{p_{j-1},p_{j-1} } - \frac{p_{j-1}}{\Omega} }\right) & \text{for an activator}    
\end{cases}  , 
\end{align}
here, $p_j$ and $M^2_{p_j,p_j}$ denote the mean concentration and second central moment of protein $P_j$, respectively. These represent Hill functions for a repressor or activator, with a Hill coefficient of two and a dissociation constant of one (see \ref{A.0}). In our case, the analysed system can be described exactly up to the second central moment because it involves only first-order reactions and functional kinetic parameters (Hill functions) \cite{Exact}. We accounted for protein diffusion, in contrast to mRNA, which does not diffuse because it remains confined within the cells. The ODEs for the mean concentrations are:
\begin{subequations} \label{20}
\begin{align} 
    \frac{\partial m_j^r}{ \partial t}=& \delta( \gamma_j^2 H_j - m_j^r), \\
    \frac{\partial p_j^r}{ \partial t}=&  m_j^r - p_j^r + \frac{p_j^{r+h}+p_j^{r-h} - 2p_j^{r}}{h^2} , 
\end{align} 
\end{subequations}
and for the second central moments
\begin{subequations} \label{21}
{\footnotesize
\begin{align} 
    \frac{\partial M^2_{m_j^r,m_j^r}}{ \partial t}=& \frac{\delta}{\Omega} \left( \gamma_j^2 H_j^r +  m_j \right) - 2 \delta M^2_{m_j^r,m_j^r}, \\
    \frac{\partial M^2_{m_j^r,p_j^r}}{ \partial t}=& M^2_{m_j^r,m_j^r} - (\delta + 1) M^2_{m_j^r,p_j^r } + \frac{M^2_{m_j^r,p_j^{r+h} }+ M^2_{m_j^r,p_j^{r-h} }-2M^2_{m_j^r,p_j^r }}{h^2} , \\
    \frac{\partial M^2_{p_j^r,p_j^r}}{ \partial t}= &\frac{1}{\Omega} \left( m_j^r +  p_j^r \right)+ 2 M^2_{m_j^r,p_j^r} - 2 M^2_{p_j^r,p_j^r} + \frac{1}{\Omega} \left( \frac{p_j^{r+h}+p_j^{r-h} + 2p_j^r}{h^2} \right) \nonumber \\
    &+ 2 \frac{M^2_{p_j^r,p_j^{r+h} }+ M^2_{p_j^r,p_j^{r-h} }-2M^2_{p_j^r,p_j^r }}{h^2} . 
\end{align}}
\end{subequations}
Eqs. \eqref{20}–\eqref{21} are a set of ODEs that provide an accurate description of the system dynamics in 1D where proteins can move to their first neighbors. The stability analysis is performed in the Fourier space, thus
\begin{subequations} 
\begin{align}
    \mathcal{F} \left[ \frac{p_j^{r+h}+p_j^{r-h} - 2p_j^{r}}{h^2} \right]=& - \left( \frac{sin\left( K \frac{h}{2} \right)}{\frac{h}{2}}\right)^2 p_j^K \nonumber \\
    =&-f_1(K,h) p_j^K,  \\
    \mathcal{F} \left[\frac{p_j^{r+h}+p_j^{r-h} + 2p_j^r }{h^2}\right]=&  \left( \frac{cos\left( K \frac{h}{2} \right)}{\frac{h}{2}}\right)^2 p_j^K \nonumber \\
    &= f_2(K,h) p_j^K.
\end{align}
\end{subequations}
The limit $\lim_{h \to 0} f_1(K,h) = K^2$, and $\lim_{h \to 0} f_2(K,h) \to \delta(0)$ corresponds to the continuous-space case \cite{Lot}.
Thus Eqs. \eqref{20} and \eqref{21} in Fourier space are
\begin{subequations} \label{24}
\begin{align}
    \frac{\partial m_j^K}{ \partial t}=& \delta( \gamma_j^2 \hat{H}_j - m_j^K),  \\
    \frac{\partial p_j^K}{ \partial t}=&  m_j^K -(1+f_1(K,h)) p_j^K , 
\end{align}
\end{subequations}
and, 
\begin{subequations} \label{25}
\begin{align}
    \frac{\partial M^2_{m_j^K,m_j^K}}{ \partial t}=& \frac{\delta}{\Omega} \left( \gamma_j^2 \hat{H}_j +  m_j \right) - 2 \delta M^2_{m_j^K,m_j^K}, \\
    \frac{\partial M^2_{m_j^K,p_j^K}}{ \partial t}=& M^2_{m_j^K,m_j^K} - (\delta + 1 + f_1(K,h)) M^2_{m_j^K,p_j^K } , \\
    \frac{\partial M^2_{p_j^K,p_j^K}}{ \partial t}= &\frac{1}{\Omega} \left( m_j^K + (1+ f_2(K,h)) p_j^K \right)+ 2 M^2_{m_j^K,p_j^K} \nonumber \\
    &- 2(1+ f_1(K,h)) M^2_{p_j^K,p_j^K} .  
\end{align}
\end{subequations}
The steady state of the system \eqref{20}-\eqref{21} without diffusion satisfies,
{\footnotesize
\begin{align}
    M^2_{m_j,m_j,ss}=& \frac{m_{j,ss}}{\Omega}, & 
     M^2_{m_j,p_j,ss}=& \frac{m_{j,ss}}{\Omega} \left( \frac{1}{\delta + 1} \right) , & M^2_{p_j,p_j,ss}=& \frac{p_{j,ss}}{\Omega}\left( 1+ \frac{1}{\delta + 1}\right).  \label{26}
\end{align}}
From these results, we can conclude that mRNA follows a Poisson distribution in the stationary state, whereas proteins deviate from a Poisson distribution. For the mean concentrations of proteins, we have 
\begin{align}
    p_{j,ss}=  m_{j,ss}. \label{27}
\end{align}
We also need to solve the following equations
\begin{align}
    m_{j,ss} = \begin{cases}        
    \gamma_j^2 \left(  \frac{1}{1+ m^2_{j-1,ss} + \frac{m_{j-1,ss}}{\Omega}\left( \frac{1}{\delta + 1}\right) }\right) & \text{for a repressor } \\  
 \gamma_j^2 \left(  \frac{m^2_{j-1,ss} + \frac{m_{j-1,ss}}{\Omega}\left( \frac{1}{\delta + 1}\right)}{1+ m^2_{j-1,ss} + \frac{m_{j-1,ss}}{\Omega}\left( \frac{1}{\delta + 1}\right) }\right) & \text{for an activator} 
\end{cases}. \label{28}
\end{align}
The linearization of the system will also involve the derivative of the Hill function evaluated at the stationary state,

{\footnotesize
\begin{align}
    \Xi_j= & \left. \frac{\partial  H_j}{ \partial p_{j-1}} \right|_{ss} = \left(  \frac{2 m_{j-1,ss} - \frac{1}{\Omega}}{ \left( 1+ m^2_{j-1,ss} + \frac{m_{j-1,ss}}{\Omega}\left( \frac{1}{\delta + 1}\right) \right)^2 }\right) \times\begin{cases}         
  -1 & \text{for a repressor } \\  
  +1 & \text{for an activator}    
\end{cases},  \label{cases}
\end{align}} 

Notice that to have a positive value between the parentheses in $\Xi_i$ it is necessary to satisfy the following condition: $m_{j, ss} > \frac{1}{2\Omega}$. We now follow a similar analysis to that in the previous subsection \ref{section3.2}. The eigenvalues of the matrix $\mathbf{E}$ in Eq. \eqref{matrix}, where $\Xi_j$ is taken from \eqref{cases} are:   
\begin{align}
    \lambda_{j}=& L_p e^{I(2j-1) \pi/n}, L_p=  \left| \prod_{j=1}^{n} \gamma_j^2 \Xi_j \right|^{\frac{1}{n}},  \label{32}
\end{align}
where $j \in (1,2,...,n)$. 

Following the approach of \cite{Harat, Hori}, control theory was employed to evaluate the system's stability. Therefore, we obtained the system's transfer functions: 

\begin{subequations} 
{\tiny
\begin{align}
    g_1(s)=&  \frac{\delta}{(\delta + s)(1 + f_1(K,h) + s)},  \\
    g_2(s)=& \frac{ \delta \left[ \delta (2 + f_1(K,h) + f_2(K,h) + s) + (1 + f_1(K,h) + s)(4 + f_1(K,h) + f_2(K,h) + s) \right] }{ (\delta + s)(1 + f_1(K,h) + s)(1 + \delta + f_1(K,h) + s)(2 + 2f_1(K,h) + s) }. 
\end{align}}
\end{subequations}

\begin{figure*}[h!t]
    \centering
    \includegraphics[width=0.3\linewidth]{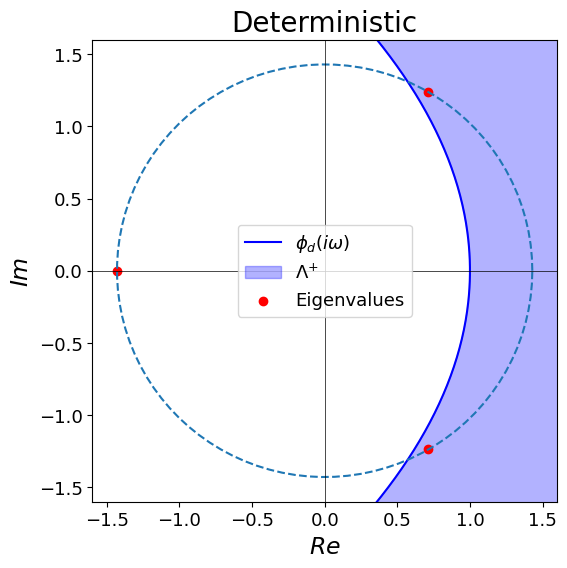}
    \includegraphics[width=0.3\linewidth]{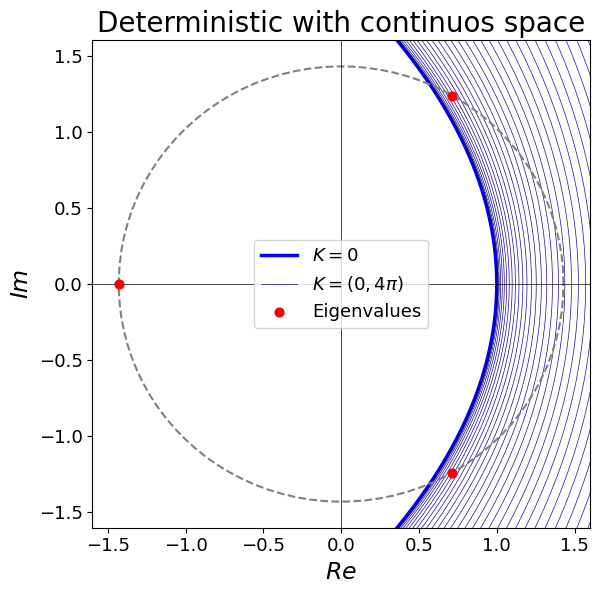}
    \includegraphics[width=0.3\linewidth]{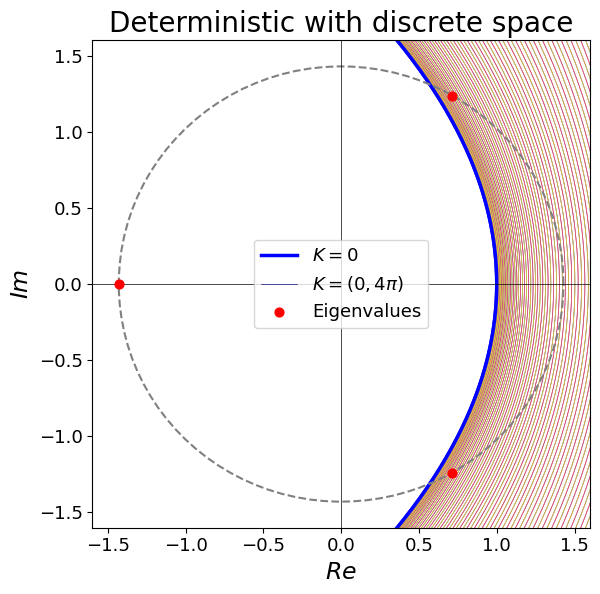}
    \caption{\textbf{Region of stability.} In this figure, we illustrate the stability criteria of Proposition 2. The first panel corresponds to the case without diffusion, where the system is unstable, and then, there are oscillations. In the other two panels, instability arises, leading to the coexistence of patterns and oscillations (Turing–Hopf instability). For each case, we analyzed both continuous and discrete space representations, and both approaches yielded the same prediction.}
    \label{fig.4}
\end{figure*}

$g_1(s)$ is associated with the equations describing the mean concentrations and thus is similar to that already obtained from the deterministic description. However, in this stochastic approach, an additional transfer function, $g_2(s)$, arises, which is associated with the second central moments. The case without diffusion is recovered when $f_1(K,h)=f_2(K,h)=0$. 

From these transfer functions, we construct the overall transfer function of the system by following a procedure similar to that used in the deterministic analysis. Thus, the system's stability is determined by the poles of  
\begin{align}
    \phi(s)= & \frac{1}{g_1(s)} \left( \prod_{j=1}^n \left( 1 + \frac{1}{\Omega} \frac{g_2(s)} {g_1(s)} \frac{\xi_j}{\Xi_j}\right) \right)^{-\frac{1}{n}}.
\end{align}

With these elements, we now state the following proposition, which summarizes the system's stability conditions under fluctuations and diffusion: \\

\textbf{Proposition 5.}  Consider a cyclic gene network with dynamics given by (\ref{24}) and (\ref{25}). Then:  

\begin{itemize}
    \item[i)] If at least one of the eigenvalues $\lambda$, exist within the domain $\Lambda^+$, defined as $\Lambda^+ = \phi(\mathbb{C}_+)$ ($\phi(s)$ is evaluated on the domain of $\mathbb{C}_+$), then the system is unstable.

    \item[ii)] Otherwise, the system is stable.
\end{itemize}

 In the limit where the system size increases and diffusion is neglected, the result reduces to the one obtained in \cite{Hori}. It is worth noticing that in this kind of system \cite{Harat, Hori, Oscillations}, instability implies a limit cycle. Moreover, Definition 3 allows us to determine whether the system exhibits Turing or Turing–Hopf instability. In the next section \ref{section4}, we analyze a particular system to investigate its dynamic behavior. 

\begin{figure}[h!t]
    \centering
\includegraphics[width=0.35\linewidth]{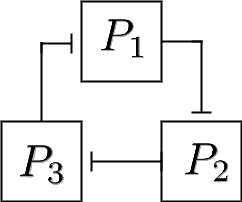}
    \caption{\textbf{Repressilator.}  This is a cyclic gene regulatory network with three modules, in which the protein represses mRNA synthesis in the next module, and the system has negative feedback.  }
    \label{fig.3}
\end{figure}

\begin{figure*}[h!t]
    \centering
    \includegraphics[width=0.41\linewidth]{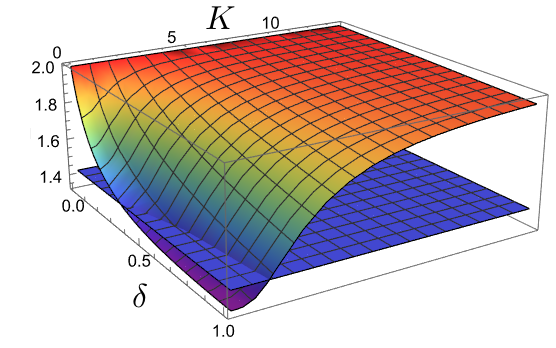}
    \includegraphics[width=0.45\linewidth]{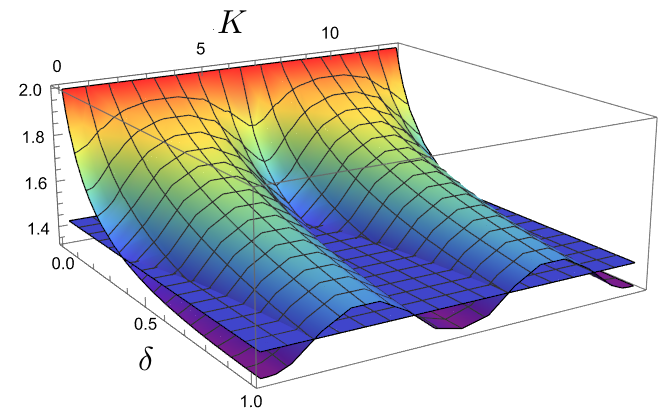}
    \caption{\textbf{Stability in discrete and continuous space.} In this figure, we plot the stability criteria from Proposition 3. The left panel corresponds to continuous space, whereas the right panel shows the discrete case ($h=1$). In this system, more instability modes appear in the discrete space because a larger number of values of $W(3,Q)$ lie below the reference level $L=1.429$. The parameter $\delta$, which represents the ratio between the degradation rates of mRNA and proteins.    }
    \label{fig.5}
\end{figure*}

\begin{figure*} [h!t]
  \begin{subfigure}{\linewidth}
\includegraphics[width=.3\textwidth]{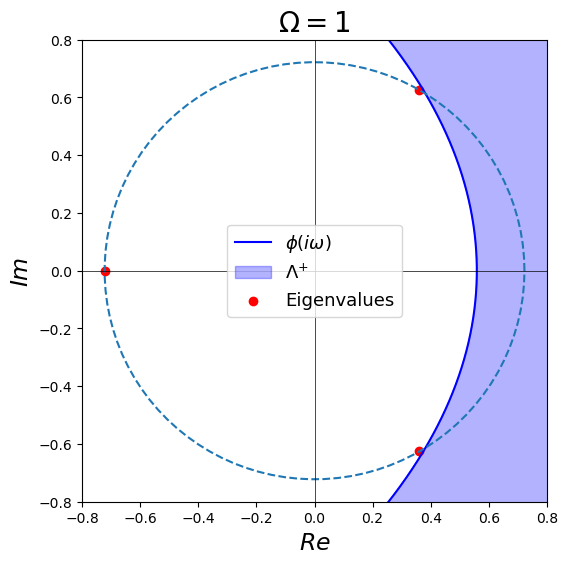}\hfill
\includegraphics[width=.3\textwidth]{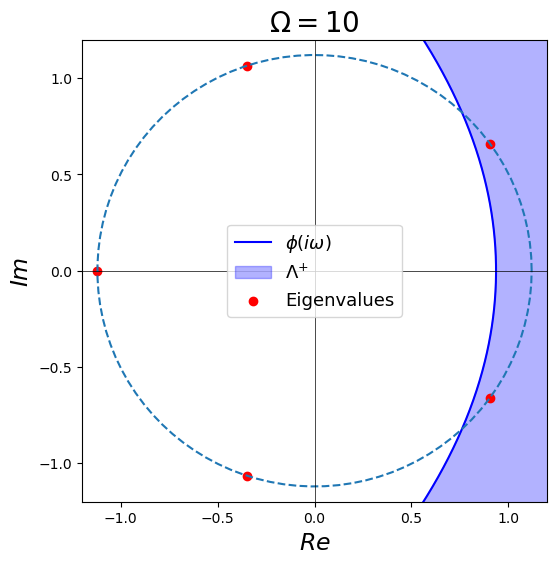}\hfill
\includegraphics[width=.3\textwidth]{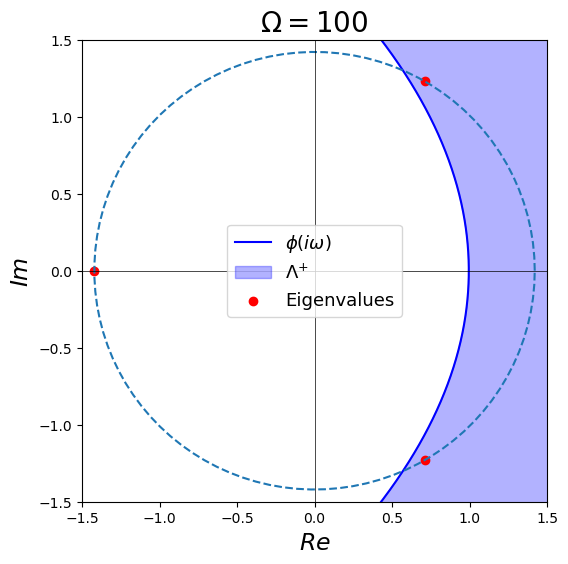}\hfill
  \end{subfigure}\par\medskip
  \begin{subfigure}{\linewidth}
  \includegraphics[width=.3\textwidth]{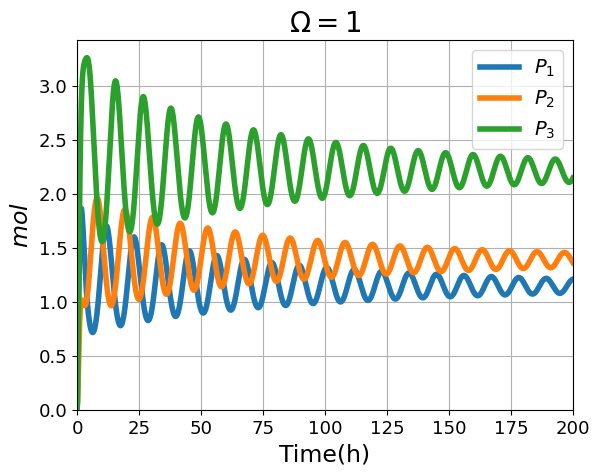}\hfill
  \includegraphics[width=.3\textwidth]{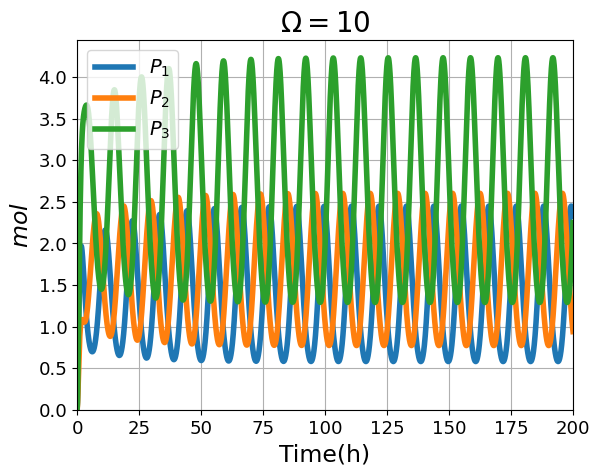}\hfill
\includegraphics[width=.3\textwidth]{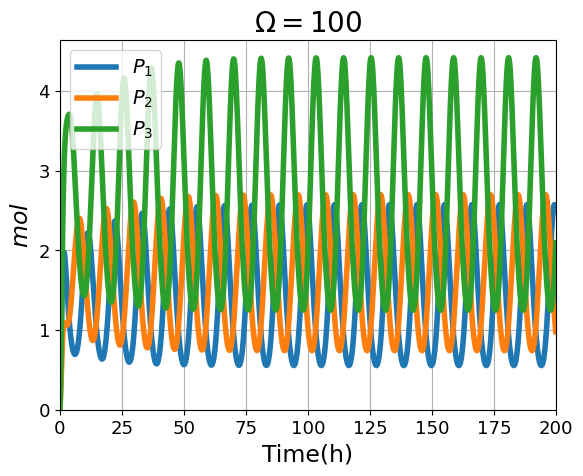}\hfill
  \end{subfigure}
\caption{\textbf{Criteria of instability without diffusion.} In this figure, we illustrate how decreasing $\Omega$ changes the stability of the system. When the mean concentration predicted instability, the dynamics exhibited oscillations consistent with the graphical criteria. The first row shows the graphical criteria for the mean concentrations and time evolution of the system, and the third row presents the second central moment and its temporal behavior. For $\Omega=1$, the system is stable, and the dynamics correspond to underdamped oscillations converging to a stationary state. We used the initial conditions of Table \ref{tab:2}. } 
\label{fig.6}
\end{figure*}

\section{Repressilator on 1D} \label{section4}

To analyze the repressilator (Fig. \ref{fig.3}), we used $n=3$, $\bm{\gamma}^2=(8,4,8)$, and $\delta=1$. For the voxel size of the space discretization, we used $h=1$, and different values of $\Omega$ were tested to explore the impact of size on the system's dynamics. The values of the stationary mean concentrations of mRNA are listed in Table \ref{tabla1}. Note that as $\Omega$ (the size of the system) decreases, the stationary values decrease. The values of the other variables were obtained using Eqs. \eqref{26} and \eqref{27} by substituting the mRNA value for the stochastic case.

\begin{table}[htbp]
\centering
\begin{tabular}{|p{2cm}|c|c|c|}
\hline
\textbf{Variable} & \textbf{Values} & $\Omega$ \\
\hline
$\mathbf{m}_{ss}$ & $(1.145, 1.388, 2.210)$  & 1 \\
$\mathbf{m}_{ss}$ & $(1.237, 1.543, 2.313)$  & 10 \\
$\mathbf{m}_{ss}$ & $(1.248, 1.560, 2.324)$  & 100 \\
$\mathbf{m}_{ss}$ & $(1.249, 1.562, 2.325)$  & $\infty$ \\
\hline
\end{tabular}
\caption{\textbf{Stationary concentrations of the mRNA.} Stationary mean concentrations of mRNA obtained from Eq. \eqref{27} for different values of $\Omega$. The last row corresponds to the deterministic case calculated using Eq. \eqref{12}. }
\label{tabla1}
\end{table}

\begin{figure*} [h!t]
  \begin{subfigure}{\linewidth}
\includegraphics[width=.3\textwidth]{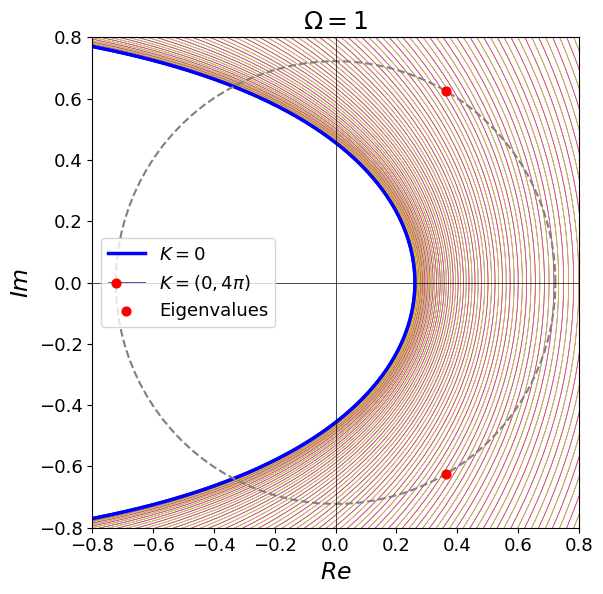}\hfill
\includegraphics[width=.3\textwidth]{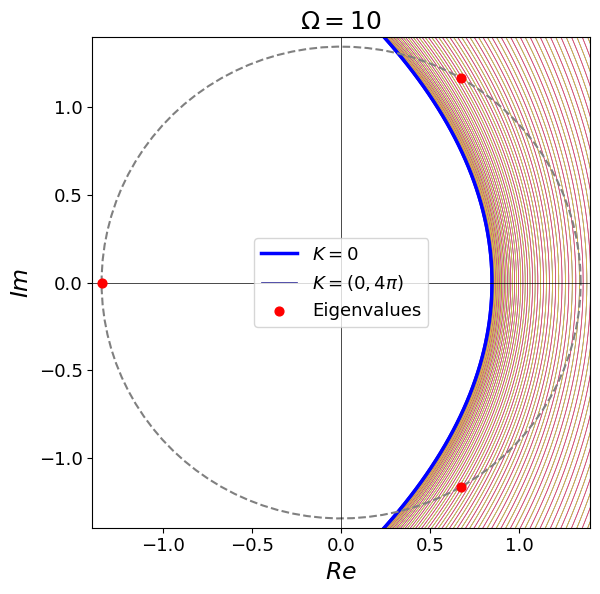}\hfill
\includegraphics[width=.3\textwidth]{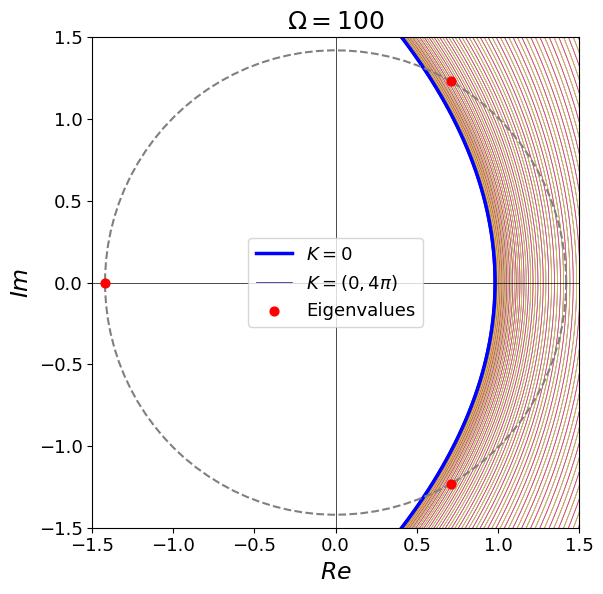}\hfill
  \end{subfigure}\par\medskip
  \begin{subfigure}{\linewidth}
  \includegraphics[width=.3\textwidth]{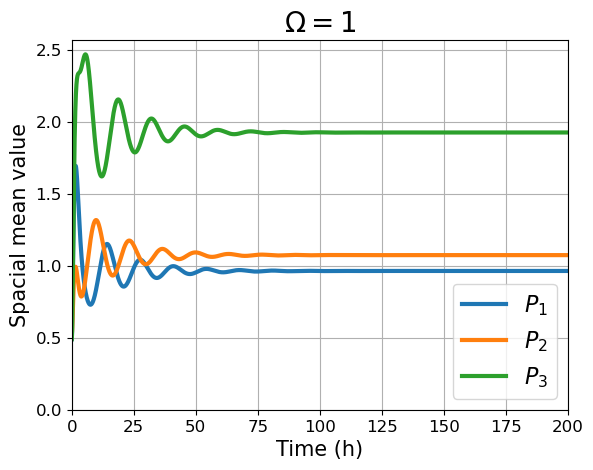}\hfill
\includegraphics[width=.3\textwidth]{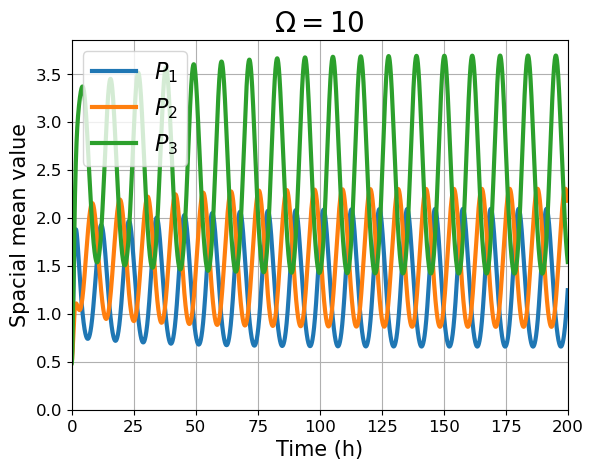}\hfill
\includegraphics[width=.3\textwidth]{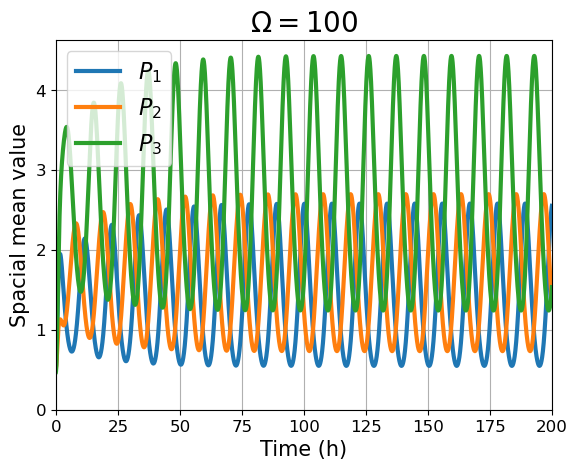}\hfill
  \end{subfigure}\par\medskip
\caption{\textbf{Criteria of stability with diffusion on 1D.} In this figure, we analyze the stability of the system for different values of $\Omega$ in the presence of diffusion. When the mean concentration predicts instability, the dynamics exhibit Turing–Hopf instability, which is in agreement with the behavior of the mean spatial concentration. The first row displays the graphical criteria for the mean concentrations and the corresponding time evolution of the mean spatial concentration, whereas the third row shows the second central moment and its temporal behavior. For $\Omega=1$, the system is stable in the absence of diffusion; however, when diffusion is included, the system becomes unstable, resulting in a Turing instability. We used random initial conditions.   } 
\label{fig.7}
\end{figure*}

\subsection{Deterministic}

In the present study, we conduct an analysis of the repressilator within a deterministic and diffusional framework. This particular gene regulatory network was originally synthesised experimentally by \cite{Elowitz}. Although previous investigations \cite{Hori, Hanna} have scrutinised this system and established the criteria for instability (oscillations), those evaluations were performed in the absence of fluctuations and diffusion.

In Fig. \ref{fig.4}, we illustrate the stability criteria of Proposition 2. The first panel corresponds to the case without diffusion, where instability is observed, and then, there are oscillations. In the other two panels, instability arises, leading to the coexistence of patterns and oscillations (Turing–Hopf instability). For each case, we analyzed both continuous and discrete space representations, and both approaches yielded the same prediction.

This naturally raises the question: if both cases predict the presence of Turing–Hopf instability, what are the differences between them? To address this, we consider the more general model given by Proposition 3 for both the continuous and discrete spaces. The results are shown in Fig. \ref{fig.5}, where we plot the function $W(n=3,Q)$ for different values of $\delta$ and $K$, which are spatial modes. The left panel shows the graphical condition for the continuous space, while the right panel corresponds to the discrete space ($h=1$). Comparing this with the value obtained from the repressilator for $L(=1.429)$, in the discrete case, a larger number of unstable modes appeared than in the continuous case. Then, different patterns can appear. 

Another important observation from Fig. \ref{fig.5} is that the case $K=0$ coincides with the situation of no diffusion. Thus, if a system is initially stable (or unstable), the inclusion of diffusion does not alter this stability property; it remains stable (or unstable). This provides a graphical illustration of Proposition 4.

\subsection{Stochastic}

Here, we analyze the repressilator considering only intrinsic fluctuations and diffusion in a discrete space,  where there are three modules, each protein represses the synthesis of the next, as shown in Fig. \ref{fig.3}, and its dynamics are described by Eqs. \eqref{24} and \eqref{25}. We analyze the stability of the system with Proposition 5.   

First, we analyzed the system without diffusion, and the corresponding results are shown in Fig. \ref{fig.6}. These results reveal that as the system size $\Omega$ decreases, the system becomes stable, and the oscillations vanish. In other words, increasing fluctuations attenuated the oscillatory behavior. An additional observation is that when $\Omega$ decreases, the dynamics are dominated by fluctuations; however, in this regime, oscillations cannot arise, as confirmed by the behavior of both the mean protein concentration and its second central moment.  

When diffusion is included (see Fig. \ref{fig.7}), the decrease in $\Omega$ stabilizes the mean protein dynamics according to the graphical criterion. Nevertheless, for the second central moment, although the mean spatial dynamics converge to a stationary state, the graphical criterion predicts the onset of an instability induced by spatial effects, namely, a stochastic Turing instability \cite{Biancalani}. To explore this phenomenon in greater detail, we extended the analysis of the model to two spatial dimensions in the next section \ref{section5}.

\section{Repressilator on 2D} \label{section5}

\begin{figure*} [h!t]
  \begin{subfigure}{\linewidth}
\includegraphics[width=.3\textwidth]{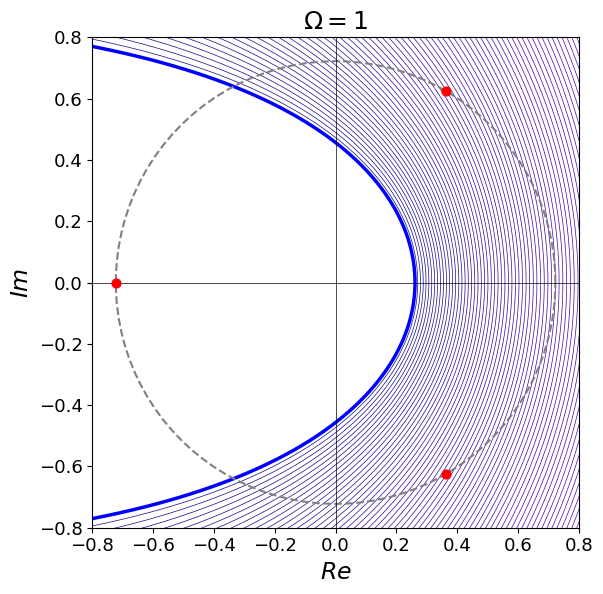}\hfill
\includegraphics[width=.3\textwidth]{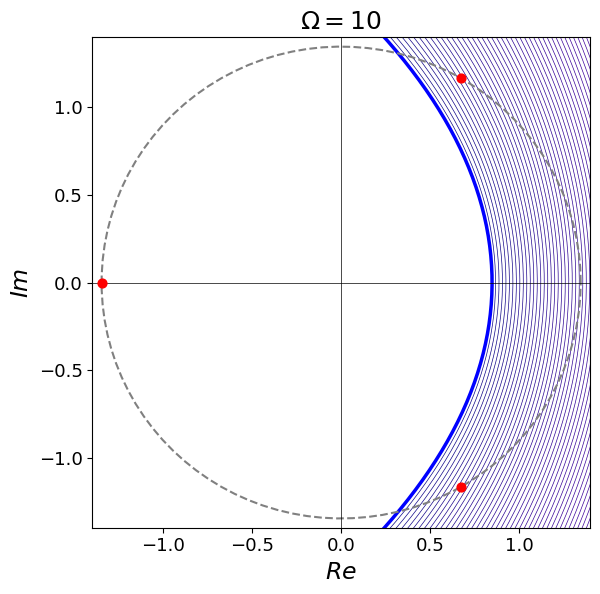}\hfill
\includegraphics[width=.3\textwidth]{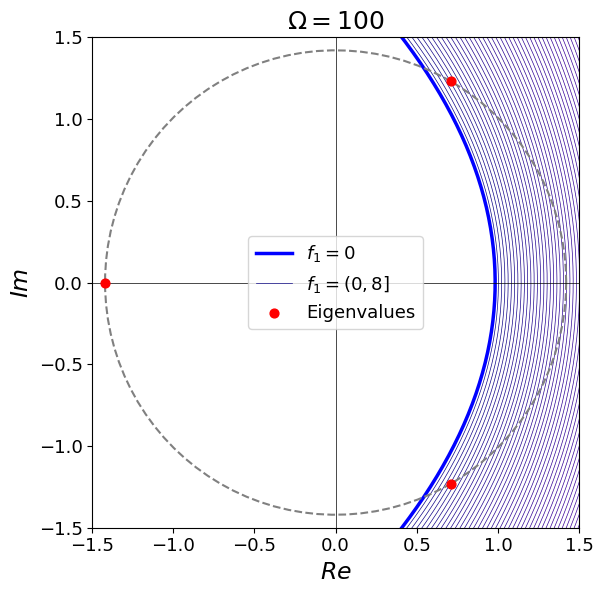}\hfill
  \end{subfigure}\par\medskip
  \begin{subfigure}{\linewidth}
  \includegraphics[width=.3\textwidth]{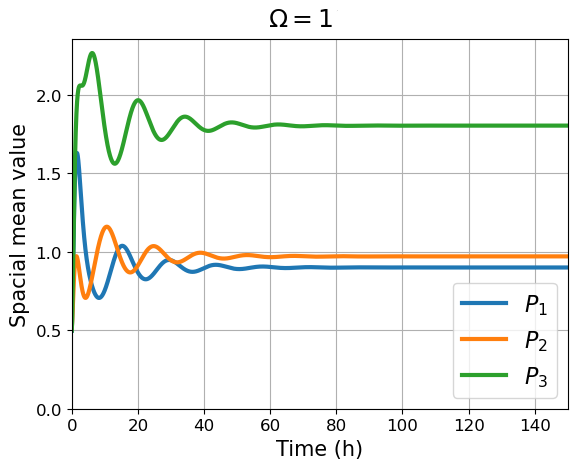}\hfill
\includegraphics[width=.3\textwidth]{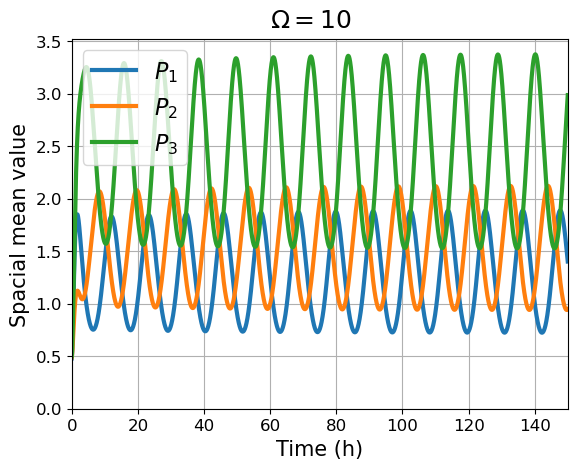}\hfill
\includegraphics[width=.3\textwidth]{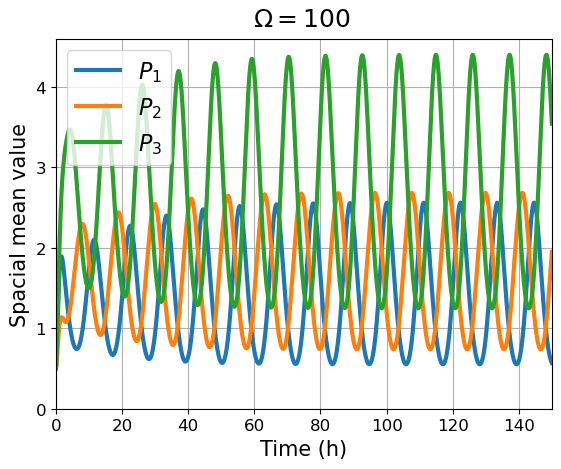}\hfill
  \end{subfigure}\par\medskip
\caption{\textbf{Criteria of stability with diffusion on 2D.}  In this figure, we analyze the stability of the system for different values of $\Omega$ in the presence of diffusion on 2D. When the mean concentration predicts instability, the dynamics exhibit Turing–Hopf instability, which is in agreement with the behavior of the mean spatial concentration. The first row displays the graphical criteria for the mean concentrations and the corresponding time evolution of the mean spatial concentration, whereas the third row shows the second central moment and its temporal behavior. For $\Omega=1$, the system is stable in the absence of diffusion; however, when diffusion is included, the system becomes unstable, resulting in a stochastic Turing instability. We used random initial conditions. } 
\label{fig.8}
\end{figure*}

\begin{figure*} [h!t]
    \centering
    \includegraphics[width=0.3\linewidth]{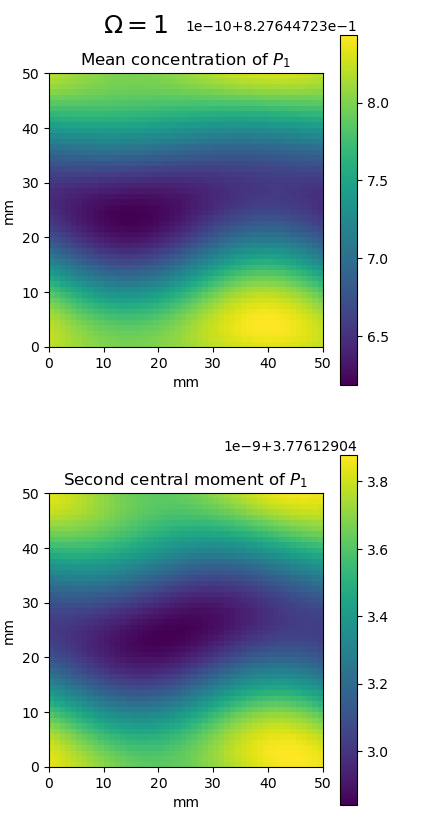}
    \includegraphics[width=0.3\linewidth]{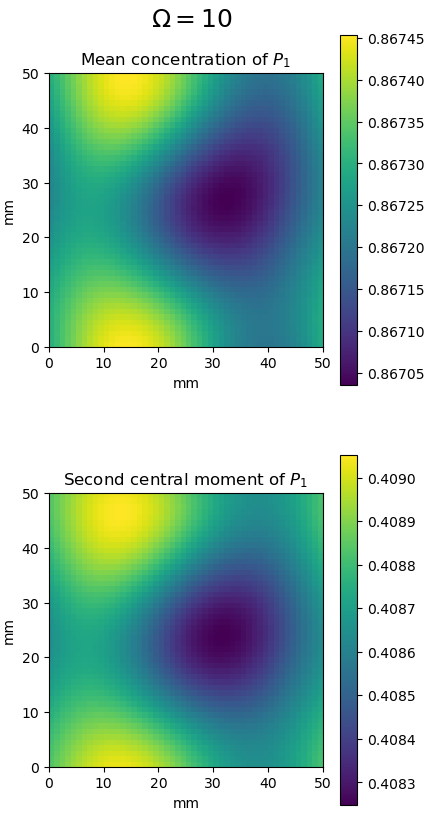}
    \includegraphics[width=0.3\linewidth]{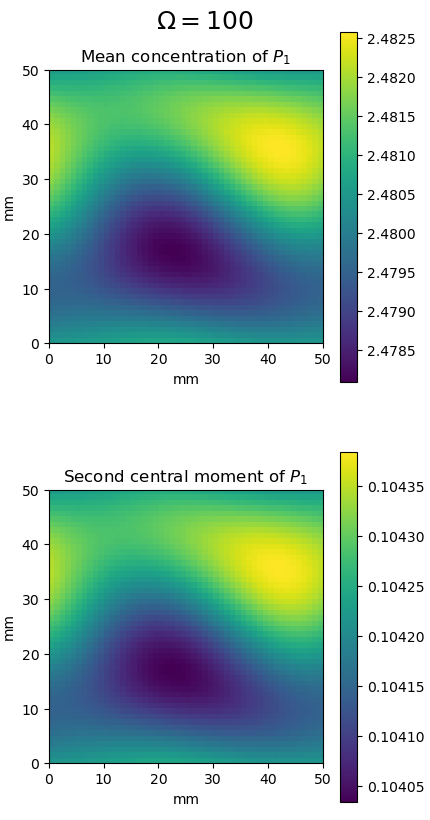}
    \caption{\textbf{2D Patterns on Repressilator.} In this figure, we present the spatial patterns of protein $P_1$ for both the mean concentration and the second central moment at time $t=200$, for different values of $\Omega$. In all cases, patterns emerge; however, the differences in their range of values remain very small. We used random initial conditions.}
    \label{fig.9}
\end{figure*}

In 1D, the stochastic system predicts that as $\Omega$ decreases, instability persists; however, in this case, the instabilities arise from the spatial dimension, Turing instability, and create a spatial pattern. To visualize this, we observed the results of the analysis in two spatial dimensions. First, to determine how this affects the system, we need to use the following functions:
\begin{subequations} 
\begin{align}
    f_1(K_1,K_2,h)=& \left( \frac{sin\left( K_1 \frac{h}{2} \right)}{\frac{h}{2}}\right)^2 + \left( \frac{sin\left( K_2 \frac{h}{2} \right)}{\frac{h}{2}}\right)^2,  \\
    f_2(K_1,K_2,h)=& \left( \frac{cos\left( K_1 \frac{h}{2} \right)}{\frac{h}{2}}\right)^2 + \left( \frac{cos\left( K_2 \frac{h}{2} \right)}{\frac{h}{2}}\right)^2,
\end{align}
\end{subequations}
where $K_1$ and $K_2$ are the modes of the pattern in each dimension. There is a relation between these two functions
\begin{align}
    f_2(K_1,K_2,h) = \frac{8}{h^2}- f_1(K_1,K_2,h).
\end{align}
For a better visualization of this in the graphical criteria, we observe that the function $f_1(K_1,K_2,h)$ has values in the interval $(0,8/h^2]$, so, for the graphical criteria, we evaluate it in this range for Proposition 5. Using the same values as in the one-dimensional case, the results are shown in Fig. \ref{fig.8}. We can see that, similar to the one-dimensional case, when $\Omega=1$ and diffusion is present, the system becomes unstable, a Turing instability appears, and patterns form. To observe the patterns, we numerically solved the set of ODEs \eqref{24} and \eqref{25} with periodic boundaries; the results are shown in Fig. \ref{fig.9}. We can see that for the three values of $\Omega$ there are patterns, which confirms the prediction that when $\Omega=1$ patterns occur, although the range of parameter values that produce patterns is very small. These patterns can be related to stochastic Turing patterns because they are produced by the second central moment, which is related to fluctuations \cite{Biancalani}. The emergence of a Turing instability in a reaction--diffusion system with equal diffusion coefficients is often regarded as a far-from-equilibrium phenomenon \cite{Ninomiya}.

The deterministic limit is recovered as $\Omega \to \infty$. For $\Omega=100$, the system dynamics closely resemble the deterministic behavior since fluctuations are very small; therefore, this case is comparable to the deterministic analysis.

\section{Results and Conclusions} \label{section6}

In this study, we extended a stability criterion to gene regulatory networks with a large number of components and diffusion, both in a deterministic framework and in a stochastic framework with intrinsic fluctuations.  In the deterministic case, we identified two main behaviors: if the system is stable (unstable) without diffusion, it remains stable (unstable) when diffusion is included. In the absence of diffusion, stability determines the presence or absence of oscillations, corresponding to a Hopf instability. However, when diffusion is present, the system may exhibit a Turing–Hopf instability. Moreover, when the spatial domain is discretized, additional modes emerge, allowing a wider range of patterns to be formed.  

In the stochastic case, if the system is stable without diffusion and the system size $\Omega$ is small, fluctuations dominate the dynamics and can induce a Turing instability, which we interpret as stochastic Turing instability. Additionally, under the assumption that all variables diffuse at the same rate, we obtained that Turing instabilities can still emerge even if the diffusion parameters are equal.  

Overall, the framework developed in this study provides a systematic method for analyzing the stability of systems with many interacting variables and diffusion effects. This simplifies the task of determining whether a system is stable or unstable and, consequently, predicting the occurrence of Turing and Turing–Hopf instabilities. This can then be applied to a broad range of biological systems that have many variables and fluctuations, both intrinsic and extrinsic.

\section*{Acknowledgments}

Manuel E. Hernández-García acknowledges the financial support of SECITHI through the program "Becas Nacionales 2023".

\section*{Declarations}

The authors declare no conflicts of interest in relation to the publication of this article. 

All data generated or analyzed in this study are included in this article.

\appendix

\section{Hill Function} \label{A.0}

Here, we derive the Hill function that is used in the principal text; this derivation is based on \cite{Exact, Decimal}. For this, we have the following reactions 
\begin{eqnarray}
    R+2L \stackbin[k_{-}]{k_{+}}{\rightleftarrows} RL_2, \nonumber \\
    0 \stackbin[k_1]{k_2}{\rightleftarrows} L, \label{c.0}
\end{eqnarray}
The first reaction is the binding of $2$ ligands $L$ to receptor $R$ in a reversible process to form complex $RL_2$. The stoichiometric coefficients and the stoichiometric matrix are, 
\begin{align}
 \alpha_{ij}&= \begin{pmatrix}
		2& 1 & 0 \\
            0& 0 & 1  \\
            0& 0 & 0  \\
            1& 0 & 0
	\end{pmatrix} ,    &
 \beta_{ij}&=  \begin{pmatrix}
		0& 0 & 1 \\
            2& 1 & 0  \\
            1& 0 & 0  \\
            0& 0 & 0 
	\end{pmatrix}, \nonumber \\
 \Gamma_{ij}&= \begin{pmatrix}
		-2 & 2 & 1 & -1 \\
		-1 & 1 & 0 & 0 \\
		1  & -1 & 0 & 0 
	\end{pmatrix}. \label{c.1}
\end{align}
Let $L, R, S$ be the number of molecules of $L, R, RL_2$ respectively. Thus, the propensity rates are
\begin{align}
    a_1&= k_{+} R \frac{L!}{(L-2)!} \frac{1}{\Omega^{3}}, & a_2=& k_{-} S\frac{1}{\Omega},     \label{c.2}
\end{align}
there is a conservative quantity $R+S=R_0$, the number of initial receptors, then the previous propensities are reduced to 
\begin{align}
    a_1&= k_{+} (R_0-S) \frac{L!}{(L-2)!} \frac{1}{\Omega^{3}}, & a_2=& k_{-} S\frac{1}{\Omega},     \label{c.3}
\end{align}
and 
\begin{align}
     \Gamma'_{ij}&= \begin{pmatrix}
		-2 & 2 & 1 & -1 \\
		1  & -1 & 0 & 0 
	\end{pmatrix},  
\end{align}
if we suppose that the first reactions in (\ref{c.0}) are in the stationary state, $S$ and $L$ are independent, then the central moments between $S$ and $L$ are zero, then we get,
\begin{align}
    \frac{\partial s}{\partial t}=0 = -k_{-} s + k_{+}(r_0 - s) \frac{1}{\Omega^{2}} \left \langle   \frac{L!}{(L-2)!}  \right \rangle,
\end{align}
where $K^2= \frac{k_{-}}{k_{+}}$,  $r_0=R_0/\Omega$, $s=\braket{S}/\Omega$ and $r=\braket{R}/\Omega$ are the mean concentrations of  $S$ and $R$, from this we get 
\begin{align}
    s= r_0 \frac{\frac{1}{\Omega^{2}}\left \langle   \frac{L!}{(L-2)!}  \right \rangle}{K^2+ \frac{1}{ \Omega^{2}}\left \langle   \frac{L!}{(L-2)!}  \right \rangle}.
\end{align}
If we define the Hill function as follows and substitute $r+s=r_0$ and the value of $s$, we have
\begin{align}
    H=\frac{s}{r+s}= \frac{s}{r_0} = \frac{\frac{1}{\Omega^{2}}\left \langle   \frac{L!}{(L-2)!}  \right \rangle}{K^2+ \frac{1}{ \Omega^{2}}\left \langle   \frac{L!}{(L-2)!}  \right \rangle}, \label{c.5}
\end{align}
using the second-moment framework and defined $l= \braket{L}/\Omega$ and $M^2_{l, l}= \braket{(L-\braket{L})^2}/\Omega^2$, we get
\begin{align}
        H= \frac{l^2 + M^2_{l,l} - \frac{l}{\Omega}  }{{K^2} + l^2 + M^2_{l,l} - \frac{l}{\Omega} }.
\end{align}
This expression is exact because we did not make any approximations in the derivation, and it is valid when the ligands bind and unbind to the receptors very fast. Deviations from the assumption of fast transcription factor binding and unbinding may compromise the validity of the Hill function approximation. In such cases, a more accurate description would require explicitly modeling the underlying biochemical reactions from which the Hill function was derived. However, this would result in a larger system of equations, thereby increasing the complexity and length of the analysis.

The derivation presented here is for an activator, but we can perform similar derivations for repressors or only use the relation $D=1-H$. For the recovery of the deterministic case, $\Omega \rightarrow \infty $ and $M^2_{l,l}=0$.

\section{Dimensionless} \label{B}

In this section, we describe how to reduce the number of parameters required to describe a system. In this case, we focus only on the ODEs for the mean concentrations because this procedure is very similar when we use all ODEs. 
\begin{subequations}
\begin{align}
    \frac{\partial m_i^r}{ \partial t}=& \gamma_{1,i}(p_{i-1}^r) - \delta_1 m_i^r, \\
    \frac{\partial p_i^r}{ \partial t}=&  \gamma_{2,i} m_i^r - \delta_2 p_i^r + D \frac{p_i^{r+h} + p_i^{r-h} - 2p_i^{r}}{h^2} ,   \\
    \gamma_{1,i}(p_{i-1})=& \gamma_{1,i}^* \left( \frac{K_0^2}{K_0^2 + p_{i-1}^2 + M^2_{p_{i-1},p_{i-1}}-\frac{p_{i-1}}{\Omega}}\right),
\end{align}
\end{subequations}
we observe that to describe the system we need $n$ parameters $\gamma_{1,i}$,  $n$ parameters $\gamma_{2,i}$, a parameter $\delta_1$, a parameter $\delta_2$, a parameter $K_0$ and a parameter $\Omega$, that is in total $2n+5$ parameters, to reduce this we introduce new variables, and substitute $m_i^r \rightarrow  m_{0,i} m_i^r$, $p_i^r \rightarrow p_{0,i} p_i^r$, $M^2_{p_{i},p_{i}} \rightarrow p_{0,i}^2 M^2_{p_{i},p_{i}}$, $\Omega \rightarrow \Omega_0 \Omega$, $h \rightarrow h_0 h$ and $t \rightarrow \tau t$, and simplify we get
\begin{align}
    \frac{\partial m_i^r}{ \partial t}=& \frac{\tau}{m_{0,i}}  \gamma_{1,i}(p_{i-1}^r) - \tau \delta_1 m_i^r, \nonumber \\
    \frac{\partial p_i^r}{ \partial t}=&  \frac{\tau m_{0,i}}{p_{0,i}} \gamma_{2,i} m_i^r - \tau \delta_2 p_i^r + \frac{\tau D}{h_0^2} \frac{p_i^{r+h} + p_i^{r-h} - 2p_i^{r}}{h^2}  ,  \nonumber \\
    \gamma_{1,i}(p_{i-1})=& \gamma_{1,i}^* \left( \frac{K_0^2}{K_0^2 + p_{0,i-1}^2 p_{i-1}^2 + p_{0,i-1}^2 M^2_{p_{i-1},p_{i-1}}- \frac{p_{0,i-1}}{\Omega_0} \frac{p_{i-1}}{\Omega}}\right), \nonumber
\end{align}
then we define $\tau= \frac{1}{\delta_2}$, $\delta=\frac{\delta_1}{\delta_2}$, $p_{0,i}= K_0$, $m_{0,i}= \frac{K_0 \delta_2}{\gamma_{2,i}}$, $\Omega_0 = \frac{1}{K_0}$, $\gamma_i^2= \frac{\gamma_{1,i}^* \gamma_{2,i}}{K \delta_1 \delta_2}$, $h_0^2= \tau D$, then 
\begin{subequations}
\begin{align}
    \frac{\partial m_i^r}{ \partial t}=& \delta( \gamma_i^*(p_{i-1}^r) -  m_i^r) , \\
    \frac{\partial p_i^r}{ \partial t}=&  m_i^r -  p_i^r + \frac{p_i^{r+h} + p_i^{r-h} - 2p_i^{r}}{h^2} ,   \\
     \gamma_i^*=& \gamma_i^2 \left( \frac{1}{1+  p_{i-1}^2 + M^2_{p_{i-1},p_{i-1}}- \frac{p_{i-1}}{K_0 \Omega}}\right),
\end{align}
\end{subequations}
from these equations, we can see that only $n$ parameters $\gamma^2_i$, a parameter $\delta$, a parameter $\Omega$, a parameter $K_0$ and a parameter $h$, that is, $n+4$ parameters, are required. But we chose $h=1$ and $K_0=1$  to reduce the number of parameters to $n+2$.

\section{Proof of Proposition 4 \label{C}}

Here, we present the proof of Proposition 4. First, we present the following:  
let $Q_0$ denote the value of $Q$ in the absence of diffusion. 

Starting with the definition of $Q^2$, we have
\begin{align}
    Q^2 &= \frac{4\delta\,(1+ f_1(K,h))}{\big(\delta + 1 + f_1(K,h)\big)^2} \nonumber \\
        &= \frac{4\delta}{(\delta +1)^2}  \frac{(1 + f_1(K,h))(\delta + 1)^2}{\big(\delta + 1 + f_1(K,h)\big)^2} \nonumber \\
        &= Q_0^2  \frac{1 + f_1(K,h)}{\left(1 + \tfrac{f_1(K,h)}{\delta + 1}\right)^2}.
\end{align}

Next, we calculate the limits of this function with respect to $f_1$:
\begin{subequations}
\begin{align}
    \lim_{f_1\to 0} \frac{1 + f_1(K,h)}{\left(1 + \tfrac{f_1(K,h)}{\delta + 1}\right)^2} &= 1,  \\
    \lim_{f_1\to \infty} \frac{1 + f_1(K,h)}{\left(1 + \tfrac{f_1(K,h)}{\delta + 1}\right)^2} &= 0. 
\end{align}
\end{subequations}
To determine whether this function attains a maximum in the interval, we computed its derivative.  It was found that the extremum occurs at $f_1=\delta-1$, if $0 < \delta \leq 1$, this maximum corresponds to a negative value, whereas $f_1$ only admits non-negative values. Hence,
\begin{align}
    \frac{1 + f_1(K,h)}{\left(1 + \tfrac{f_1(K,h)}{\delta + 1}\right)^2} \leq 1,
\end{align}

\begin{table*}
\centering
\begin{tabular}{ccc}
\hline
\hline
\textbf{Variable} & \textbf{Description} & \textbf{Value}\\
\hline
$m_j(0)$& Initial mean concentration of $m_j$. & $0$ $mol$ \\
${p}_j(0)$& Initial mean concentration of $p_j$. & $0$ $mol$ \\
$M^2_{m_j,m_j}(0)$& Initial second central moment of concentration {mRNA}. & 0 $mol^2$ \\
$M^2_{m_j,p_j}(0)$& Initial second central moment of concentration {mRNA} and proteins. & 0 $mol^2$ \\
$M^2_{p_j,p_j}(0)$& Initial second central moment of concentration proteins. & 0 $mol^2$ \\
\hline
\end{tabular} 
\caption{\textbf{Initial conditions for the case without diffusion.} Initial conditions for the system. ($j\in (1,2,3)$) In this table, we show the initial conditions used for the gene regulatory network with negative feedback.}  \label{tab:2}
\end{table*}

and therefore
\[ Q^2 \leq Q_0^2. \]
but in the other case, if $\delta>1$, we evaluate in the maximum value, then 
\begin{align}
    \frac{1 + f_1(K,h)}{\left(1 + \tfrac{f_1(K,h)}{\delta + 1}\right)^2}= \frac{\delta}{4} \left(1 +\frac{1}{\delta} \right)^2 > 1,
\end{align}
and therefore, for some $K$ and $h$ the next relation exist, 
\[ Q_0^2 < Q^2. \]
It is important to mention that in the biological systems the value of $\delta>1$ because the degradation of the ARm is more fast that the degradation of proteins.  

\medskip

Now, we show the relation between $W$ and $W_0$ ($W_0$ denote the value of $W$ in the absence of diffusion) for the two scenarios.  To do this, we demonstrate that $W$ is a monotonically decreasing function of $Q^2$.  
Differentiating, we obtain
\begin{align}
   \partial_{Q^2} W
   &= \frac{2}{\sin^2\left( \frac{\pi}{n} \right)}
   \left( - \frac{ -\cos\left( \frac{\pi}{n} \right) 
   + \sqrt{\cos^2\left( \frac{\pi}{n}\right) + Q^2 \sin^2\left( \frac{\pi}{n} \right) } }{Q^4 } \right. \nonumber \\
   &+ \left. \frac{1}{2 Q^2} \frac{\sin^2\left( \frac{\pi}{n} \right)}
   {\sqrt{\cos^2\left( \frac{\pi}{n}\right) + Q^2 \sin^2\left( \frac{\pi}{n} \right) }} \right) \nonumber \\
   &= - \frac{1}{Q^4 \sin^2\left( \frac{\pi}{n} \right) 
   \sqrt{\cos^2\left( \frac{\pi}{n}\right) + Q^2 \sin^2\left( \tfrac{\pi}{n} \right) }} \times \nonumber \\
   &\left( \sqrt{\cos^2\left( \frac{\pi}{n}\right) + Q^2 \sin^2\left( \frac{\pi}{n} \right)} 
   - \cos\left( \frac{\pi}{n}\right) \right)^2 .
\end{align}

This shows that $\partial_{Q^2}W \leq 0$, that is, the function $W$ monotonically decreases with respect to $Q^2$.  
For the case $\delta \leq 1$, $Q^2 \leq Q_0^2$, it follows that
\[
W_0 \leq W.
\]

We now explore the four cases of Definition 3.

\medskip

\textbf{Case 1.} In the absence of diffusion, the system is stable.  
From Proposition 3, we have $L^D \leq W_0$.  
When diffusion is added, the system remains stable; therefore, $L^D \leq W$.  
Both conditions are consistent, $L^D\leq W_0 \leq W$; therefore, this scenario is possible.

\medskip

\textbf{Case 2.} In the absence of diffusion, the system is stable.  
From Proposition 3, we have $L^D \leq W_0$.  
When diffusion is added, the system becomes unstable; therefore, $L^D > W$.  
However, since $W_0 \leq W$, this implies simultaneously that $L^D \leq W_0$ and $L^D > W \geq W_0$, which is a contradiction.  
Therefore, this scenario is not possible.

\medskip

\textbf{Case 3.} In the absence of diffusion, the system is unstable.  
From Proposition 3, we have $L^D > W_0$.  
When diffusion is added, the system becomes stable; thus, $L^D \leq W$.  
However, since $W_0 \leq W$, this would imply simultaneously $L^D > W_0$ and $L^D \leq W_0 \leq W$, which is a contradiction.  
Therefore, this scenario is not possible.

\medskip

\textbf{Case 4.} In the absence of diffusion, the system is unstable.  
From Proposition 3, we have $L^D > W_0$.  
When diffusion is added, the system remains unstable; therefore, $L^D > W$. This implied that $L^D > W \geq W_0$.
Both conditions are consistent; therefore, this scenario is possible.

\medskip

For the case in which $\delta>1$, we have $Q_0^2 < Q^2$, then
\[
W < W_0.
\]
We now explore the four cases of Definition 3.

\medskip

\textbf{Case 1.} In the absence of diffusion, the system is stable.  
From Proposition 3, we have $L^D \leq W_0$.  
When diffusion is added, the system remains stable; therefore, $L^D \leq W$.  This implies  that $L^D \leq W_0$ and $L^D \leq W < W_0$, which is possible only when $L^D < W_0$.


\medskip

\textbf{Case 2.} In the absence of diffusion, the system is stable.  
From Proposition 3, we have $L^D \leq W_0$.  
When diffusion is added, the system becomes unstable; therefore, $L^D > W$.  
This implies that $W_0 \geq L^D > W$, however, since $W_0 > W$, this scenario is possible.

\medskip

\textbf{Case 3.} In the absence of diffusion, the system is unstable.  
From Proposition 3, we have $L^D > W_0$.  
When diffusion is added, the system becomes stable; thus, $L^D \leq W$.  
However, since $W < W_0$, this would imply simultaneously  $W \geq L^D > W_0$ and $L^D > W_0>W$ and, which is a contradiction.  
Therefore, this scenario is not possible.

\medskip

\textbf{Case 4.} In the absence of diffusion, the system is unstable.  
From Proposition 3, we have $L^D > W_0$.  
When diffusion is added, the system remains unstable; therefore, $L^D > W$. This implied that $L^D > W_0 > W$.
Both conditions are consistent; therefore, this scenario is possible.

\medskip
This completes the proof of Proposition 4.

\section{Parameters and initial conditions}

Table \ref{tab:2} lists the initial conditions of the represilador without diffusion presented in this study.

{

}

\end{document}